\documentclass[a4paper,12pt]{article}
\usepackage{soul}
\usepackage[usenames,dvipsnames]{color}
\usepackage{jheppub}
\usepackage{esvect}
\usepackage{amsmath, amssymb, slashed, epsf, color, graphicx, latexsym}
\usepackage{epsfig}
\usepackage{graphics}

\newcommand{\BR}{{\bar{R}}}
\newcommand{\Bnabla}{{\bar{\nabla}}}

\begin{document}
	
%	\preprint{get one}
	\title{The large $D$ Membrane Paradigm For Einstein-Gauss-Bonnet Gravity}
	\author{Arunabha Saha}
	\affiliation[a]{University of Geneva, 24 quai Ernest-Ansermet, 1211 Geneve 4, Switzerland}
	\affiliation[b]{Department of Theoretical Physics, Tata Institute of Fundamental Research, Mumbai, India-400005}
	\emailAdd{arunabha.saha@unige.ch}
	
	\abstract{We find the equations of motion of membranes dual to the black holes in Einstein-Gauss-Bonnet (EGB) gravity to leading order in $1/D$ in the large $D$ regime.~We also find the metric solutions to the EGB equations to first subleading order in $1/D$ in terms of membrane variables. We propose a world volume stress tensor for the membrane whose conservation equations are equivalent to the leading order membrane equations. We work out the light quasi-normal mode spectrum of static black holes in EGB gravity from the linearised fluctuations of static, round membranes. Also, the effective equations for stationary black holes and the spectrum of linearised spectrum about black string configurations has been obtained using the membrane equation for EGB gravity.~All our results are worked out to linear order in the Gauss-Bonnet parameter. }

%opening
\maketitle

%\tableofcontents
\section{Introduction}
In the limit of spacetime dimensions ($D$) going to infinity, the late time dynamics (at time scales of order $\mathcal{O}(1)$) of black holes consists of only a finite number of degrees of freedom. The primary evidence for this was presented in \cite{Emparan:2014aba}, where it was shown that the spectrum of light quasi-normal modes (with frequency of order $\mathcal{O}(1)$) consists of two scalar and one vector mode at each value of angular momentum number of the background $SO(D-1)$ isometry group. There are an infinitely large number of heavy quasi-normal modes of frequency of the order $\mathcal{O}(D)$ which die off in time intervals of the order of $\mathcal{O}(1/D)$. Motivated by this the authors of \cite{Bhattacharyya:2015dva} derived a set of effective non-linear equations determining this late time dynamics in terms of variables of a membrane propagating in flat spacetime. This duality between black hole and membranes propagating in flat spacetimes is known as the large $D$ membrane paradigm. %Motivated by this in \cite{Bhattacharyya:2015fdk}, an effective equation determining the full non-linear gravitational dynamics of these light modes was obtained in terms of a set of equations determining the dynamics of a dual membrane. 
Details of this duality in Einstein-Maxwel system \cite{Bhattacharyya:2015fdk} and  in presence of a cosmological constant \cite{Bhattacharyya:2017hpj, Bhattacharyya:2018szu} have also been worked out. In \cite{Dandekar:2016fvw}, a systematic algorithm was proposed to work out the membrane dynamics to  any order in $1/D$ and explicit computations to find the first subleading correction to membrane equations and next to leading order corrections in metric were presented\footnote{An equivalent description of the effective equations describing the large $D$ dynamics of black holes has been worked out in many different situations: $1)$ static and stationary cases \cite{Emparan:2015hwa, Suzuki:2015iha} $2)$ Black string dynamics \cite{Suzuki:2015axa, Emparan:2015gva}, $3)$ charged black holes \cite{Tanabe:2015isb}. In \cite{Dandekar:2016jrp, Dandekar:2017aiv}, the equivalence between the two apparently different approaches to find the effective equations was established. Other interesting developments along these lines can be found in \cite{Emparan:2016sjk, Herzog:2016hob, Rozali:2017bll, Rozali:2016yhw, Herzog:2017qwp}}. 

Since the large $D$ limit has been very useful in understanding many aspects of black hole physics in two-derivative Einstein gravity, it is natural to apply this methodology to study black hole dynamics in higher derivative theories of gravity.  Study of the effective theory of spacetimes with horizons has already been initiated for the simplest higher derivative theories of gravity namely the Einstein-Gauss-Bonnet gravity. In \cite{Chen:2015fuf} the quasi-normal mode spectrum of static spherically symmetric black holes in EGB gravity in the large $D$ limit has been worked out. It was found that when the Gauss-Bonnet (GB) terms can be treated as perturbations over the Einstein-Hilbert terms in the equations of motion, the same number of characteristic light modes mentioned above are still present and the frequencies of the light modes are corrected by the presence of the GB terms. The large $D$ effective equations of stationary black holes for EGB gravity has also been obtained in \cite{Chen:2017wpf}. The large $D$  effective equations of dynamics within a region of width $\mathcal{O}(1/D)$ about static black holes in presence of a GB term was obtained in \cite{Chen:2017hwm}.Further a detailed analysis of linearised modes about Static black holes was also carried out and it was shown that the spectrum matches the gravitational computation of QNMs in \cite{Chen:2015fuf}.Similar effective equations of dynamics about black strings with spatial derivatives of the order $\mathcal{O}(\sqrt{D})$ has been worked out in \cite{Chen:2017rxa} and the Hydro-elastic complementarity study of \cite{Emparan:2016sjk} and the study of turbulent regime of fluid dynamics in large $D$ initiated in \cite{Rozali:2017bll} was also extended to Einstein-Gauss-Bonnet (EGB) gravity in \cite{Chen:2018nbh}. 

In this paper we follow a complementary path of deriving the membrane equations dual to dynamical black holes in the large $D$ limit. The equations obtained in this paper are on the same footing as those presented in \cite{Bhattacharyya:2015dva, Dandekar:2016fvw} etc. They capture the $O(D^0)$ derivative dynamics in the leading large $D$ limit for objects with horizon. The membrane equations developed here are local and hence, one need not specify the global topology of the object with horizon to start with. One can nevertheless apply the membrane equations to specific situations where the topology and other properties of the horizon are given in some detail and derive specific dynamical equations adapted to those. The membrane equations dual to two derivative gravity \cite{Bhattacharyya:2015dva, Dandekar:2016fvw} were used to study the effective equations for static and stationary horizon configurations in \cite{Dandekar:2017aiv, Mandlik:2018wnw}. It was shown in \cite{Dandekar:2016jrp} that the effective equations about black branes with derivatives of the order of $\mathcal{O}(\sqrt{D})$ \cite{Emparan:2015gva} can be obtained by suitable redefinitions of the membrane variables.

More precisely in this paper we will work out the details of the large $D$ membrane paradigm for black holes in EGB gravity to leading order in $1/D$. We work to linear order in the GB parameter ($\beta$) only. The degrees of freedom of the dual membrane are same as those for the membranes dual to black holes in two derivative Einstein gravity, namely its shape and a time-like velocity vector field defined on the membrane. We find the equation of motion of the membrane dual to the black holes in EGB gravity to leading order in a $1/D$ expansion. It is given by
\begin{eqnarray}
&&\nabla.u=\mathcal{O}(\frac{1}{D})\quad \text{and,}\nonumber\\
&&\mathcal{P}_A^M\left(\frac{\nabla^2 u_{M}}{\mathcal{K}}+u^{B}K_{BM}-u.\nabla u_M\left(1+\beta \frac{\mathcal{K}^2}{(D-3)^2}\right)-\frac{\nabla_M\mathcal{K}}{\mathcal{K}}\left(1-\beta\frac{\mathcal{K}^2}{(D-3)^2}\right)\right)=\mathcal{O}\left(\frac{1}{D}\right)\nonumber\\
\end{eqnarray}
Here $u$ is unit normalised time like velocity vector field on the membrane and $\mathcal{K}$ is the trace of extrinsic curvature $K_{AB}$ of the membrane embedded in flat spacetime.~We also work out the metric of the dynamical black hole which solves the EGB equation to first subleading order in $1/D$ in terms of membrane variables(see equation \eqref{geom_metric} and \eqref{geom_metric_exp}). Also, we write down an expression for the world volume stress tensor of the dual membrane given by 
 \begin{eqnarray}
 8\pi T_{MN}&=&\frac{\mathcal{K}}{2}\Bigg(1+\frac{\beta\mathcal{K}^2}{D^2}\Bigg)u_M u_N+\Bigg(1-\frac{\beta\mathcal{K}^2}{D^2}\Bigg)\frac{K_{MN}}{2}-\frac{\nabla_M u_N+\nabla_N u_M}{2}-\left(u_M V_N+u_N V_M\right)\nonumber\\
 \text{where}&&\nonumber\\
 V_M&=& -\frac{1}{2}\left(1-\frac{\beta\mathcal{K}^2}{D^2}\right)\frac{\nabla_M\mathcal{K}}{\mathcal{K}}+\frac{\beta\mathcal{K}^2}{D^2}u^A K_{AM}-\frac{\beta\mathcal{K}^2}{2D^2}u\cdot \nabla u_M+\left(1+\frac{\beta\mathcal{K}^2}{D^2}\right)\frac{\nabla^2u_M}{\mathcal{K}}\nonumber\\
 \end{eqnarray}
 The equations of conservation of this stress tensor in the world volume of the membrane embedded in flat spacetime is equivalent to the membrane equations written above.~We also find the spectrum of light quasi-normal modes of static black holes in EGB gravity in the large $D$ limit from the linearised fluctuations of the membrane about a static spherical configuration. The frequency of the scalar modes is given by
 \begin{equation}
 w_s=(1-\beta)\left(-(l-1)\pm \sqrt{l-1}\right)+\mathcal{O}\left(\frac{1}{D}\right),
 \end{equation}
 and the frequency of the velocity mode is given by
 \begin{equation}
 w_v=-(1-\beta)(l-1)+\mathcal{O}\left(\frac{1}{D}\right).
 \end{equation}
 Here, $l$ denotes the angular  momentum number of the modes for the spherical harmonics in the respective tensor sectors with respect to the $SO(D-1)$ isometry direction of the background. The frequency of the scalar mode matches with the linearised spectrum of the effective equations obtained in \cite{Chen:2017hwm}.
 
 We will also evaluate the membrane equations obtained here  for stationary membranes to obtain the following effective equation for the trace of the extrinsic curvature of the membrane $\mathcal{K}$ in terms of the red-shift factor $\gamma$. 
\begin{equation}
\mathcal{K}=\frac{\gamma D}{\alpha_0}+\beta \left(\frac{\gamma \alpha_1 D}{\alpha_0^2}+\frac{\gamma^3 D}{\alpha_0^3}\right).
\end{equation}	
Here, $\alpha_0$ and $\alpha_1$ are constants determining the temperature of the dual black hole. It will also be shown that these effective equations are in agreement with the equations obtained in \cite{Chen:2017hwm} to linear order in GB parameter. 

In addition we also analyse the spectrum of linearised fluctuations about a membrane dual to black strings \cite{Dandekar:2016jrp} using the modified membrane equations and find that the spectrum matches with the corresponding answers in \cite{Chen:2017rxa} under a rescaling of spatial derivatives.
 %In this paper we lay the gorundwork 
  
%Till now most of the work in studying large $D$ black holes have been confined to studying them in two-derivative Einstein gravity (with or without charge). Due to the simple picture the large $D$ membrane paradigm gives to the complex dynamics involving black holes, it is not very difficult to see that this methodology has a great potential to shed more light on many aspects of black hole physics in higher derivative theories of gravity. The most pressing of these questions being the status of the second law of black hole thermodynamics in higher derivative theories of gravity. Some initial steps in studying large $D$ black holes in these theories have already been initiated in \cite {Chen:2015fuf, Chen:2016fuy, Chen:2017hwm, Chen:2017rxa}. In these papers the quasinormal modes of these black holes have been worked out in large $D$ limit and also the effective equations for static black holes and black strings in these theory have been worked out. 
\section{EGB gravity: lagrangian, equations of motion and large $D$ scaling of the GB parameter}
\subsection{The lagrangian}
 The Einstein-Gauss Bonnet (EGB) gravity is the simplest in a series of consistent\footnote{ By consistent we mean there are no -super-luminal modes for linearised fluctuations about flat spacetime. For fluctuations about other spacetimes, some constraint on the parameters can be imposed to make sure there are no ghost propagating modes} higher derivative terms that can be added to the Einstein-Hilbert term.~They are generically called the Lovelock theories of gravity.~The lagrangian for Lovelock theories is given by
\begin{eqnarray}\label{Lovelock action}
&&\mathcal{L}=\sqrt{-g}\sum_{n=1}\alpha_nR^{(n)}\nonumber\\
\text{where}\quad &&R^{(n)}=\frac{1}{2^n}\delta ^{\mu_1\nu_1...\mu_n\nu_m}_{\alpha_1\beta_1...\alpha_n\beta_n}\prod_{r=1}^nR^{\alpha_r\beta_r}_{~~~~\mu_r\nu_r},\nonumber\\
&&\delta ^{\mu_1\nu_1...\mu_n\nu_m}_{\alpha_1\beta_1...\alpha_n\beta_n}=n! \delta^{\mu_1}_{[\alpha_1}\delta^{\nu_1}_{\beta_1}...\delta^{\mu_n}_{\alpha_n}\delta^{\nu_n}_{\beta_n]}\nonumber\\
\text{and}\quad &&R^{\alpha_r\beta_r}_{~~~~\mu_r\nu_r} \quad \text{is the Riemann tensor}.
\end{eqnarray}
The square bracket above denotes anti-symmetrisation in the indices without any numerical factor. The leading term in the series (i.e. when $n=1$) is the Einstein-Hilbert term. 
We will be working with only the first two terms in this series which together form the EGB gravity lagrangian. The explicit form of the EGB lagrangian is given by\footnote{ Here without loss of generality we can set $\alpha_1=1$ and $\alpha_2=\alpha$}.
\begin{equation}
\mathcal{L}=\sqrt{-g}\left(R+\alpha \left(R^2+R_{\mu\nu\beta\gamma}R^{\mu\nu\beta\gamma}-4R_{\mu\nu}R^{\mu\nu}\right)\right).
\end{equation}
We will be thinking of $\alpha$ as a perturbatively small parameter and in this paper we will be working to linear order in $\alpha$ only. From the structure of the lagrangian \eqref{Lovelock action} we see that the parameters $\alpha_n$ count the number of derivatives on the metric for the particular term in the lagrangian. Hence, effects linear in $\alpha$ come from $4$ derivative terms in the action and so on. If we want to consider effects which are quadratic or higher order in $\alpha$ (which are in principal $8$ derivative effects), we feel that it is necessary to also take into account other terms in the lagrangian which explicitly contain higher number of derivatives on the metric. Hence to unambiguously keep track of four derivative terms in the lagrangian we only work to  linear order in $\alpha$.
%\footnote{In a more complete theory of gravity which generically contains all the higher derivative terms in the lagrangian the EGB lagrangian can be thought of as an effective theory which makes sense only upto $4$ derivative orders and hence makes it inconsistent to consider higher order in one of the couplings but ignoring the similarly relevant higher derivative terms present in the theory. We assume this to be the case and proceed with our study}.  
\subsection{Equation of motion and large $D$ scaling of the GB parameter}
 The variation of the EGB action w.r.t. the metric gives rise to the following equation of motion 
\begin{eqnarray}\label{original_GB_eqn}
E_{AB}=&& R_{AB}-\frac{1}{2}g_{AB}R-\alpha\Bigg(\frac{1}{2}g_{AB}\left(R_{EFCD}R^{EFCD}-4R_{EF}R^{EF}+R^2\right)\nonumber\\
&&-2R R_{AB}+4R_{AC}R^C_B-4R^{CD}R_{CABD}-2R_{ACDF}R_B^{CDF}\Bigg)
\end{eqnarray}

Static black hole metrics in $D$ dimensions contain  blackening factors which go like $1-\frac{C}{r^{D-3}}+...$, where $r$ measures radial distance in spherical polar coordinates and $C$ is a constant. In the large $D$ limit the maximum order in $D$ that can appear in the expression of Riemann tensor comes from two $r$ derivatives acting on the blackening factor and hence $R^A_{~BCD}=\mathcal{O}(D^2)$. The Ricci tensor is obtained from contraction of two of the indices of Riemann tensor and if this contraction happens along the unit sphere part of the Schwarzschild metric then an additional factor of $D$ can arise in Ricci tensor. But we know that to zeroth order in $\alpha$ the black hole metric satisfies $R_{AB}=0$ and hence $R_{AB}=\mathcal{O}(D^2)$. Similarly, since the black hole metric is also a zero of the Ricci Scalar to zeroth order in $\alpha$, we have $R=\mathcal{O}(D^2)$. Hence, to linear order in $\alpha$ the maximum order in $D$ coming from the Gauss-Bonnet part of the equations is $\mathcal{O}(D^4)$, as it is quadratic in internal curvatures of the metric. 

 %In our analysis we will be interested  in situations which preserves a huge isometry group $SO(D-p-3)$ as $D->\infty$ where $p$ is a finite positive integer. Due to this isometry the non-zero Riemann tensor components which have two or all four legs in the isometry direction are proportional to combinations of unit sphere metric along the isometry directions. And it turns out that the Ricci tensor formed by contraction of two indices of Riemann tensor should be one order higher in $D$ than the Riemann tensor. But we see from EGB equation of motion \eqref{original_GB_eqn} that this $\mathcal{O}(D^3)$ piece must also be $\mathcal{O}(\alpha)$. Hence, this leading contribution in large $D$ cannot contribute to the EGB equation. Similar arguments hold for the Ricci scalar contribution to the EGB term. Hence, the maximum leading order contribution from the EGB terms in the large $D$ limit is of the order $\mathcal{O}(D^4)$ as all the terms there are quadratic in the intrinsic curvatures.   

%Hence in these situations taking a trace brings extra factors of $D$ because in the isometry directions \footnote{e.g. Look at earlier papers on the large $D$ membrane paradigm. At an intuitive level it is due to the fact that the trace acts as something proportional to `identity' along the huge number of isometry directions and contribute an exra factor of $D$}. Hence, naively the Ricci tensor $R_{AB}=\mathcal{O}(D^3)$. But to $\mathcal{O}(\alpha^0)$ the background Einstein equations sets $R_{AB}=R=0$. Hence, all of Riemann tensor, Ricci Tensor and Ricci Scalar mus be $\mathcal{O}(D^2)$. 

So, naively it looks like that the $\mathcal{O}(\alpha^0)$ terms in the equation \eqref{original_GB_eqn} are sub-dominant as compared to the $\mathcal{O}(\alpha)$ terms in the large $D$ limit. This makes it implausible to have  solution to the EGB equation in the large $D$ limit which are  perturbatively close to the solution of pure Einstein's equation. But we expect such solutions to exist as we consider the Lovelock theories to be corrections of Einstein-Hilbert gravity. The only plausible solution to this apparent puzzle is that $\alpha$ must be $\mathcal{O}(D^{-2})$. More specifically we will choose $\alpha$ to be given by $$\alpha=\frac{\beta}{(D-3)(D-4)},$$ where $\beta$ is an $\mathcal{O}(1)$ quantity. We will find in the next section that this specific choice helps us to write the metric of the black hole in a manner where the only $D$ dependence in the metric is in the fall off behaviour of the blackening factor as the black hole metric approaches the asymptotically flat spacetime. Written in terms  $\beta$ the EGB equation becomes
\begin{eqnarray}\label{GB_eqn}
E_{AB}=&& R_{AB}-\frac{1}{2}g_{AB}R-\frac{\beta}{(D-3)(D-4)}\Bigg(\frac{1}{2}g_{AB}\left(R_{EFCD}R^{EFCD}-4R_{EF}R^{EF}+R^2\right)\nonumber\\
&&-2R R_{AB}+4R_{AC}R^C_B-4R^{CD}R_{CABD}-2R_{ACDF}R_B^{CDF}\Bigg)
\end{eqnarray}
\section{The Static Black Hole in EGB gravity and the starting ansatz}
The static spherically symmetric Schwarzschild like black-hole solution for EGB gravity in Kerr-Schild form is given by
\begin{equation}
ds^2= -dt^2+dr^2+r^2d\Omega_{D-2}^2+(1-f(r))(dt-dr)^2
\end{equation}
where the blackening factor $f(r)$ is given by
\begin{equation}
f(r)=1+\frac{r^2}{2\beta}\left(1-\sqrt{1+\frac{4\beta r_h^{D-3}}{r^{D-1}}\left(1+\frac{\beta}{r_h^2}\right)}\right)
\end{equation}
In the above $r_h$ is the position of the horizon. Since we are interested in effects only linear in $\beta$, the relevant form of the blackening factor is
\begin{equation}\label{SC_blackhole}
f(r)=1-\left(\frac{r_h}{r}\right)^{D-3}\left(1+\frac{\beta}{r_h^2}\right)+\beta\frac{r_h^{2(D-3)}}{r^{2(D-2)}} +\mathcal{O}(\beta^2)
\end{equation}
The blackening factor can be written in a more covariant looking form as
\begin{equation}\label{leadingansatz}
f=1-\frac{1}{\psi^{D-3}}\left(1+\beta\frac{\mathcal{K}^2}{(D-2)^2}\right)+\beta\frac{\mathcal{K}^2}{(D-2)^2}\frac{1}{\psi^{2(D-2)}}
\end{equation}
where $\mathcal{K}$ is the trace of extrinsic curvature of the surface $r=r_h$, embedded in flat spacetime and is given by $\mathcal{K}=\frac{D-2}{r_h}.$ The horizon is now denoted by the surface $\psi=1$ since $\psi=\frac{r}{r_h}$.
Let us now define two new vectors: $1)$ a `velocity' vector $u=-dt$ and a vector normal to constant $\psi$ slices given by $n=dr$. The vectors are orthonormal in the sense that
\begin{equation}\label{orthonormality}
u.u=-1, \quad n.n=1\quad \text{and}\quad n.u=0
\end{equation}
where the dot products are taken w.r.t the flat spacetime. Let us also define a unique combination of these two vectors $O=n-u$. Using these variables the static black hole metric in EGB gravity can be written as
\begin{equation}\label{ansatz_metric}
ds^2=ds_{flat}^2+\left(\frac{1}{\psi^{D-3}}\left(1+\beta \frac{\mathcal{K}^2}{(D-2)^2}\right)-\beta \frac{\mathcal{K}^2}{(D-2)^2}\frac{1}{\psi^{2(D-2)}}\right)\left(O_\mu dx^\mu\right)^2
\end{equation}
%The above metric is a more covariant way of depicting the static black hole metric where $\psi$ and $u$ are constant scalar and vector respectively satisfying the properties \eqref{orthonormality}. 

In the large $D$ limit the above metric has a fast direction along $d\psi$ as it goes to flat spacetime exponentially fast outside the horizon in this direction. This can be seen by probing regions of width of the order of $\mathcal{O}(1/D)$ outside the horizon using the change of variable  $$\psi=1+\frac{R}{D}.$$ With this change of variable in the $D\rightarrow\infty$ limit the blackening factor becomes $$f(R)=1-e^{-R}\left(1+\beta \frac{\mathcal{K}^2}{(D-2)^2}\right) +\beta e^{-2R} \frac{\mathcal{K}^2}{(D-2)^2}.$$ So, there is non-trivial warping of the spacetime only up to distances of the order of $\mathcal{O}(1/D)$ outside the horizon where it decays exponentially fast to the asymptotically flat spacetime. 

One can now promote the quantities $\psi$ and $u$ to be scalar and vector functions of the spacetime coordinates with order $\mathcal{O}(1)$ derivatives. The spacetime is still flat when $\psi-1\sim \mathcal{O}(1)$ and hence it continues to solve the EGB equations there.  In the non-trivial region where $\psi-1\sim \mathcal{O}(1/D)$, the EGB equations are no longer satisfied and we will need to correct the metric \eqref{ansatz_metric} to make sure it satisfies the EGB equation. This correction can be found in a perturbative series in  $1/D$ and most of the paper will be dedicated to finding the leading order correction to the metric.  We call the metric \eqref{ansatz_metric} the `starting ansatz metric'\footnote{As a side remark we notice that the starting ansatz metric looks exactly like the staring ansatz for charged black holes in \cite{Bhattacharyya:2015fdk} if we make the replacement $$\beta \frac{\mathcal{K}^2}{(D-2)^2}\rightarrow Q^2,$$ where $`Q'$ is the charge parameter used in the paper \cite{Bhattacharyya:2015fdk}. These similarities will also persist in some aspects of the final results. } as it solves the EGB equations to the leading order approximation of the functions $\psi$ and $u$ being constants. 

The starting ansatz metric can be thought of as being dual to a membrane propagating in flat-spacetime with the shape of the membrane being given by the equation $\psi=1$. $u$ can be thought of as the time-like velocity of points on the membrane along its world volume. Given the shape and velocity of the membrane the dual black hole metric is given by \eqref{ansatz_metric}. The corrections to the metric will also be written in terms of the shape and velocity data on the membrane. 
%If instead of the constant values that the objects $\psi, u$ and $n$ take, let us now consider them to be smooth functions of the spacetime coordinates with order $\mathcal{O}(1)$ derivatives. Now around a given point in spacetime to leading order, the functions still take constant values and hence, will solve the EGB equations to leading order in $1/D$. But the metric fails to solve the EGB equations at first subleading order and so on. In this paper we will systematically correct the above metric in a $1/D$ expansion, so that the final metric solves the EGB equations to first subleading order in $1/D$. 
\section{Auxilliary conditions on the data of the membrane}
The information about the shape of the membrane dual to the black hole is contained entirely in the zeroes of the function $\psi-1$. And similarly, the data about the velocity field is contained entirely on the function $u$ defined on the surface $\psi=1$. 
%How these functions behave of the surface of the membrane has no bearing on the properties of the membrane. 
But in order to get a local expression for the black hole metric we need to define a way to lift these functions away from the membrane surface. In writing the ansatz metric \eqref{ansatz_metric}, we made a particular choice for these extensions of the $\psi=1$ surface by globally defining $\psi=\frac{r_h}{r}$ and $u=-dt$. This choice though convenient is not `covariant' and hence difficult to implement in our computations. Instead we make use of the choices used in \cite{Bhattacharyya:2015dva, Dandekar:2016fvw} and uplift the $\psi$ function such that 
\begin{equation}\label{auxilliary_condn_psi}
n.\nabla n=0.
\end{equation}
 Similarly, the $u$ function is uplifted using the equation
 \begin{equation}
 n.\nabla u=0
 \end{equation}
The covariant derivatives in the above expressions are defined w.r.t. the flat spacetime. We will call the above constraints on the $\psi$ and $u$ functions as the `auxiliary conditions'. These auxiliary conditions are not essential to defining the membrane and hence the above choice is in no way unique and  there are other valid choices for these auxiliary conditions (see e.g. \cite{Bhattacharyya:2015fdk, Bhattacharyya:2017hpj}). The difference between various choices of auxiliary conditions will appear in the final expression for the metric corrections, but the physical content remains the same. When the EGB equations are applied to the ansatz metric \eqref{ansatz_metric}, the equations are not solved beyond leading order in $1/D$. The `amount' by which they are off from being a solution will depend on the particular choice of auxiliary conditions and hence the correction that needs to be added to make them a solution also depends on the choice of the auxiliary conditions. But the final metric at a given order in $1/D$ should be the same once they are expressed in terms of the data on the membrane since, they solve the same set of equations with same set of boundary conditions. 

%So, basically the corrected metric which solves EGB equation at subleading order in $1/D$ must all be physically the same if they are evaluated using the same boundary (and/or initial condition) on the metric. This must be the case if the EGB equations are to  define a set of equations which uniquely track the `evolution' of the metric. 

\section{The isometry ansatz and the effective equations}
  Taking a large $D$ limit makes sense only when we keep some quantity fixed while taking this limit. If we are to gain insight into some aspects of finite $D$ black hole dynamics from studying the problem in the large $D$ limit the natural thing to do will be to work with configurations which preserve a large isometry $SO(D-p-3)$ as $D\rightarrow \infty$, keeping $p$ fixed. In these configurations all the non-trivial dynamics takes place in the finite number of $p+3$ directions. A very useful method to implement this is to reduce the full $D$ dimensional Einstein's equation into a set of effective equations in $p+3$ dimensional `effective spacetime'  \cite{Bhattacharyya:2015dva, Bhattacharyya:2015fdk, Dandekar:2016fvw}. 

 To do this we choose the most general metric which preserves the isometry mentioned above, namely,
\begin{equation}
ds^2=g_{\mu\nu}(x)dx^{\mu}dx^{\nu}+e^{\phi(x)}d\Omega_{d},\quad \text{where}\quad d=D-p-3.
\end{equation}
The equation \eqref{GB_eqn} reduce to a set of equation for the metric $g_{\mu\nu}$ of the effective $p+3$ dimensional space-time and the scalar field $\phi$.  Let $\mu\nu...$ denote the coordinates of the $p+3$ dimensional effective spacetime and $i,j,..$ denote the coordinates on the $d$ dimensional sphere. Also, the indices $A,B...$ denote coordinates of the full $D$ dimensional spacetime.  Due to the isometry the effective equations can only have both the indices in the $\mu\nu...$ directions or in the $i,j...$ directions The equations with both the legs in the $\mu\nu...$ directions  are given by
\begin{eqnarray}\label{tensor_eff_eqn}
&& E_{\alpha\beta}= R_{\alpha\beta}-\frac{1}{2}g_{\alpha\beta}R-\frac{\beta}{(D-3)(D-4)}\Bigg(\frac{1}{2}g_{\alpha\beta}\left(R_{ABCD}R^{ABCD}-4R_{AB}R^{AB}+R^2\right)\nonumber\\
&&~~~~~~~~~-2R R_{\alpha\beta}+4R_{\alpha C}R^C_{\beta}-4R^{CD}R_{C\alpha\beta D}-2R_{\alpha CDF}R_{\beta}^{CDF}\Bigg)\quad\text{where}\nonumber\\
&&R_{\alpha\beta}=\BR_{\alpha\beta}-\frac{d}{2}\Bnabla_{\alpha}\Bnabla_{\beta}\phi-\frac{d}{4}\Bnabla_{\alpha}\phi\Bnabla_{\beta}\phi\nonumber\\
&& R=\BR-d\Bnabla^2\phi-\frac{d(d+1)}{4}\Bnabla_{\alpha}\phi\Bnabla^{\alpha}\phi+d(d-1)e^{-\phi}\nonumber\\
&& R_{ABCD}R^{ABCD}=\BR_{\alpha\beta\gamma\delta}\BR^{\alpha\beta\gamma\delta}+4d\left(\frac{1}{2}\Bnabla_{\nu}\Bnabla_{\gamma}\phi+\frac{1}{4}\Bnabla_{\nu}\phi\Bnabla_{\gamma}\phi\right)\left(\frac{1}{2}\Bnabla^{\nu}\Bnabla^{\gamma}\phi+\frac{1}{4}\Bnabla^{\nu}\phi\Bnabla^{\gamma}\phi\right)\nonumber\\
&&~~~~~~~~~~~~~~~~~~~~~~+2e^{-2\phi}d(d-1)\left(\frac{1}{4}\Bnabla_{\mu}\phi\Bnabla^{\mu}\phi e^{\phi}-1\right)^2\nonumber\\
&&R_{AB}R^{AB}=\left(\BR_{\alpha\beta}-\frac{d}{2}\Bnabla_{\alpha}\Bnabla_{\beta}\phi-\frac{d}{4}\Bnabla_{\alpha}\phi\Bnabla_{\beta}\phi\right)\left(\BR^{\alpha\beta}-\frac{d}{2}\Bnabla^{\alpha}\Bnabla^{\beta}\phi-\frac{d}{4}\Bnabla^{\alpha}\phi\Bnabla^{\beta}\phi\right)\nonumber\\
&&~~~~~~~~~~~~~~~~+d\left((d-1)e^{-\phi}-\frac{1}{2}\Bnabla^2\phi-\frac{d}{4}\Bnabla_{\mu}\phi\Bnabla^{\mu}\phi\right)\nonumber\\
&& R_{\mu}^AR_{A\nu}=g^{\gamma\delta}\left(\BR_{\mu\gamma}-\frac{d}{2}\Bnabla_{\mu}\Bnabla_{\gamma}\phi-\frac{d}{4}\Bnabla_{\mu}\phi\Bnabla_{\gamma}\phi\right)\left(\BR_{\delta\nu}-\frac{d}{2}\Bnabla_{\delta}\Bnabla_{\nu}\phi-\frac{d}{4}\Bnabla_{\delta}\phi\Bnabla_{\nu}\phi\right)\nonumber\\
&&R^{AB}R_{A\mu\nu B}=g^{\beta\delta}\left(\BR_{\beta\gamma}-\frac{d}{2}\Bnabla_{\beta}\Bnabla_{\gamma}\phi-\frac{d}{4}\Bnabla_{\beta}\phi\Bnabla_{\gamma}\phi\right)\BR^{\gamma}_{\mu\nu\delta}\nonumber\\
&&~~~~~~~~~~~~~~~~~~+d\left((d-1)e^{-\phi}-\frac{1}{2}\Bnabla^2\phi-\frac{d}{4}\Bnabla_{\mu}\phi\Bnabla^{\mu}\phi\right)\left(\frac{1}{2}\Bnabla_{\mu}\Bnabla_{\nu}\phi+\frac{1}{4}\Bnabla_{\mu}\phi\Bnabla_{\nu}\phi\right)\nonumber\\
&&R_{\mu ABC}R_{\nu}^{ABC}=\BR_{\mu\alpha\beta\gamma}\BR_{\nu}^{\alpha\beta\gamma}+2d\left(\frac{1}{2}\Bnabla_{\mu}\Bnabla_{\eta}\phi+\frac{1}{4}\Bnabla_{\mu}\phi\Bnabla_{\eta}\phi\right)\left(\frac{1}{2}\Bnabla_{\nu}\Bnabla^{\eta}\phi+\frac{1}{4}\Bnabla_{\nu}\phi\Bnabla^{\eta}\phi\right)\nonumber\\
\end{eqnarray}
where the barred quantities are w.r.t the $p+3$ dimensional effective metric. 
The second set of equations with both legs in the $i,j...$ directions are proportional to the metric of the $d$ dimensional sphere and hence is effectively just one more equation given by
\begin{eqnarray}\label{scalar_eff_eqn}
&& E_{ij}= R_{ij}-\frac{1}{2}e^\phi \Omega_{ij}R-\frac{\beta}{(D-3)(D-4)}\Bigg(\frac{1}{2}e^\phi\Omega_{ij}\left(R_{ABCD}R^{ABCD}-4R_{AB}R^{AB}+R^2\right)\nonumber\\
&&~~~~~~~~~-2R R_{ij}+4R_{i C}R^C_{j}-4R^{CD}R_{Cij D}-2R_{i CDF}R_{i}^{CDF}\Bigg)\quad\text{where}\nonumber\\
&&R_{ij}=e^{\phi}\Omega_{ij}\left((d-1)e^{-\phi}-\frac{1}{2}\Bnabla^2\phi-\frac{d}{4}\Bnabla_{\mu}\phi\Bnabla^{\mu}\phi\right)\nonumber\\
&& R=\BR-d\Bnabla^2\phi-\frac{d(d+1)}{4}\Bnabla_{\mu}\phi\Bnabla^{\mu}\phi+d(d-1)e^{-\phi}\nonumber\\
&& R_{i}^AR_{Aj}=e^{\phi}\left((d-1)e^{-\phi}-\frac{1}{2}\Bnabla^2\phi-\frac{d}{4}\Bnabla_{\mu}\phi\Bnabla^{\mu}\phi\right)^2\Omega_{ij}\nonumber\\
&& R^{AB}R_{AijB}=\Bigg(e^{\phi}\left(\BR^{\mu\nu}-\frac{d}{2}\Bnabla^{\mu}\Bnabla^{\nu}\phi-\frac{d}{4}\Bnabla^{\mu}\phi\Bnabla^{\nu}\phi\right)\left(\frac{1}{2}\Bnabla_{\mu}\Bnabla_{\nu}\phi+\frac{1}{4}\Bnabla_{\mu}\phi\Bnabla_{\nu}\phi\right)\nonumber\\
&&~~~~~~~~~~~~~~~~~~+(d-1)\left((d-1)e^{-\phi}-\frac{1}{2}\Bnabla^2\phi-\frac{d}{4}\Bnabla_{\mu}\phi\Bnabla^{\mu}\phi\right)\left(\frac{1}{4}\Bnabla^{\mu}\Bnabla_{\mu}e^{\phi}-1\right)\Bigg)\Omega_{ij}\nonumber\\
&&R_{i ABC}R_{j}^{ABC}=2e^{\phi}\Omega_{ij}\left(\frac{1}{2}\Bnabla_{\alpha}\Bnabla_{\beta}\phi+\frac{1}{4}\Bnabla_{\alpha}\phi\Bnabla_{\beta}\phi\right)\left(\frac{1}{2}\Bnabla^{\alpha}\Bnabla^{\beta}\phi+\frac{1}{4}\Bnabla^{\alpha}\phi\Bnabla^{\beta}\phi\right)\nonumber\\
&&~~~~~~~~~~~~~~~~~~~~+2(d-1) e^{-\phi}\Omega_{ij}\left(\frac{e^{\phi}}{4}\nabla_{\alpha}\phi\nabla^{\phi}-1\right)^2
\end{eqnarray}
Total number of degrees of freedom in the metric $g_{\mu\nu}$ are $\frac{(p+3)(p+4)}{2}$ and one degree of freedom coming from the scalar $\phi$. This matches precisely with the total number of effective equations, so naively it looks like that we have a set of well defined deterministic set of equations.

\section{Zooming in to the near horizon region and the scaled coordinates}\label{patch}
Since the non-trivial physics of black holes is confined to a region of width $\mathcal{O}(1/D)$ outside the horizon, it will be convenient  to use a coordinate system which is obtained by zooming into the a small region of size $1/D$ about a point $x_0$ in this non-trivial region. This will be particularly useful to us as we will be performing most of our computations on computer algebra and working with these coordinates will make it much easier to keep track of orders of $D$ \footnote{One can as well work in the global coordinates and keep track of the orders of $D$ carefully, if one is willing to perform the computation without the help of computer algebra. This has been done to successfully work out the details of the dual membrane and the metric corrections for black holes in $AdS$ in \cite{Bhattacharyya:2017hpj, Bhattacharyya:2018szu}. Though we find it much more convenient to use computer algebra.}. The coordinates adopted to this zoomed in region is given by\footnote{We have used $D-3$ to scale the coordinates instead of $D$. We use this convention  as $D-3$ appears in the fall of behaviour of the black hole metric at asymptotic infinity. }

\begin{equation}
x^\mu=x^\mu_0+\frac{1}{D-3}~ \alpha^{\mu}_ay^a.
\end{equation}
Which implies,  $$y^a=(D-3)(x^\mu-x^\mu_0)\alpha^a_\mu.$$
But the metric in the $y^a$ coordinates still measures order $\mathcal{O}(1/D^2)$ distance for order $\mathcal{O}(1)$ coordinates distance as shown below. 
\begin{eqnarray}
&&g_{ab}=\frac{\partial x^\mu}{\partial y^a}\frac{\partial x^\nu}{\partial y^b}g_{\mu\nu}\nonumber\\
\implies&&g_{ab}=\frac{1}{(D-3)^2}\alpha^\mu_a\alpha^\nu_bg_{\mu\nu}
\end{eqnarray}
We rescale our metrics in this region so that they measure order $\mathcal{O}(1)$ distance for order $\mathcal{O}(1)$ coordinates distance in the following manner.
\begin{eqnarray}
 &&G_{ab}=(D-3)^2g_{ab}=\alpha^\mu_a\alpha^\nu_bg_{\mu\nu}\nonumber\\
&&G^{ab}=\frac{1}{(D-3)^2}g^{ab}=\alpha^a_\mu\alpha^b_\nu g^{\mu\nu}
\end{eqnarray}
Similarly, the derivatives in the $y^a$ coordinates become order $\mathcal{O}(1/D)$ 
\begin{equation}
\partial_a\phi=\frac{1}{D-3}\alpha^\mu_a\partial_\mu\phi
\end{equation}
And we define new scaled variable $\chi_a$ given by
\begin{equation}
\chi_a=(D-3)\partial_a\phi=\alpha^\mu_a\partial_\mu\phi
\end{equation}
%Therefore, the scaled metric $G_{ab}$ and unscaled metric $g_{ab}$ have the same Riemann tensor $R^a_{bcd}$ and 
%\begin{eqnarray}
%R^{(G)}_{ab}=R^{d(G)}_{adb}=R^{d(g)}_{adb}=R^{(g)}_{ab}\nonumber\\
%R^{(G)}=\frac{R^{(g)}}{(D-3)^2}
%\end{eqnarray}

The above rescalings make sure that in addition to working in the membrane region, we also work with metrics which measure $\mathcal{O}(1)$ distances and the derivatives are also $\mathcal{O}(1)$. This will help us to keep track of only those orders of $D$ which appear solely due effects from the presence of black holes and the large isometry. 

The effective equations \eqref{tensor_eff_eqn} and \eqref{scalar_eff_eqn}  expressed in terms of the scaled metric $G_{ab}$ and $\chi_a$ are given by
\begin{eqnarray}
&&T_1 - T_2 - \beta ((GB_1 - 4 GB_2 + GB_3) - 2 NT_1 + 4 NT_2 - 
4 NT_3 - 2 NT_4)=0\nonumber\\
&&\text{and}\nonumber\\
&&t_1-t_2-\beta(t_3+t_4+t_5+t_6+t_7)=0
\end{eqnarray}
where
\begin{eqnarray}
T_1&=&\BR_{ab}-\frac{\epsilon d}{2}\nabla_a\chi_b-\frac{d \epsilon^2}{4}\chi_a\chi_b\nonumber\\
T_2&=&\frac{G_{ab}}{2}\left(\BR-d\epsilon\nabla_a\chi^a-\frac{d(d+1)}{4}\epsilon^2\chi_a\chi^a+\frac{d(d-1)}{\sigma^2}\epsilon^2\right)\nonumber\\
GB_1&=&\frac{G_{ab}}{2}\Bigg(\BR_{abcd}\BR^{abcd}+4 d \epsilon^2\left(\frac{1}{2}\nabla_a\chi_b+\frac{\epsilon}{4}\chi_a\chi_b\right)\left(\frac{1}{2}\nabla^a\chi^b+\frac{\epsilon}{4}\chi^a\chi^b\right)\nonumber\\
&+&\frac{2\epsilon^4}{\sigma^4}d(d-1)\left(\frac{1}{4}\chi_a\chi^a\sigma^2-1\right)^2\Bigg)\nonumber\\
GB_2&=&\frac{G_{ab}}{2}\Bigg(\left(\BR_{ab}-\frac{d}{2}\epsilon \nabla_a\chi_b-\frac{d\epsilon^2}{4}\chi_a\chi_b\right)\left(\BR^{ab}-\frac{d}{2}\epsilon \nabla^a\chi^b-\frac{d\epsilon^2}{4}\chi^a\chi^b\right)\nonumber\\
&+&d\epsilon^4\left(\frac{d-1}{\sigma^2}-\frac{1}{2\epsilon}\nabla_a\chi^a-\frac{d}{4}\chi_a\chi^a\right)^2\Bigg)\nonumber\\
GB_3&=&\frac{G_{ab}}{2}\left(\BR-d\epsilon\nabla_a\chi^a-\frac{d(d+1)}{4}\epsilon^2\chi_a\chi^a+\frac{d(d-1)}{\sigma^2}\epsilon^2\right)^2\nonumber\\
NT_1&=&\left(\BR-d\epsilon\nabla_a\chi^a-\frac{d(d+1)}{4}\epsilon^2\chi_a\chi^a+\frac{d(d-1)}{\sigma^2}\epsilon^2\right)\left(\BR_{ab}-\frac{\epsilon d}{2}\nabla_a\chi_b-\frac{d \epsilon^2}{4}\chi_a\chi_b\right)\nonumber\\
NT_2&=&G^{cd}\left(\BR_{ac}-\frac{d\epsilon}{2}\nabla_a\chi_c-\frac{d\epsilon^2}{4}\chi_a\chi_c\right)\left(\BR_{db}-\frac{d\epsilon}{2}\nabla_d\chi_b-\frac{d\epsilon^2}{4}\chi_d\chi_b\right)\nonumber\\
NT_3&=&G^{de}\left(\BR_{ce}-\frac{d\epsilon}{2}\nabla_c\chi_e-\frac{d\epsilon^2}{4}\chi_c\chi_e\right)\BR^c_{abd}\nonumber\\
&+&\epsilon^3 d\left(\frac{d-1}{\sigma^2}-\frac{1}{2\epsilon}\nabla_a\chi^a-\frac{d}{4}\chi_a\chi^a\right)\left(\frac{1}{2}\nabla_a\chi_b+\frac{\epsilon}{4}\chi_a\chi_b\right)\nonumber\\
NT_4&=&\BR_{acde}\BR_b^{cde}+2d\epsilon^2\left(\frac{1}{2}\nabla_a\chi_e+\frac{\epsilon}{4}\chi_a\chi_e\right)\left(\frac{1}{2}\nabla_b\chi^e+\frac{\epsilon}{4}\chi_b\chi^e\right)
\end{eqnarray}
and,
\begin{eqnarray}
t_1&=&(d-1)-\frac{\sigma^2}{2\epsilon}\nabla_a\chi^a-\frac{d\sigma^2}{4}\chi_a\chi^a\nonumber\\
t_2&=&\frac{\sigma^2}{2}\Bigg(\frac{\BR}{\epsilon^2}-\frac{d}{\epsilon}\nabla_a\chi^a-\frac{d(d+1)}{4}\chi_a\chi^a+\frac{d(d-1)}{\sigma^2}\Bigg)\nonumber\\
t_3&=&\frac{1}{2}\left(gb_1-4gb_2+gb_3\right)\quad\text{where}\quad\nonumber\\
gb1&=&\sigma^2\Bigg(\frac{\BR_{abcd}\BR^{abcd}}{\epsilon^2}+4d\left(\frac{\epsilon}{4}\chi_a\chi_b+\frac{1}{2}\nabla_a\chi_b\right)\left(\frac{\epsilon}{4}\chi^a\chi^b+\frac{1}{2}\nabla^a\chi^b\right)\nonumber\\
&&+2\epsilon^2\sigma^{-4}(d-1)d\left(\frac{\chi_a\chi^a}{4}\sigma^2-1\right)^2\Bigg)\nonumber\\
gb_2&=& \sigma^2\Bigg(\frac{1}{\epsilon^2}\left(\BR_{ab}-\frac{d\epsilon}{2}\nabla_a\chi_b-\frac{d\epsilon^2}{4}\chi_a\chi_b\right)\left(\BR^{ab}-\frac{d\epsilon}{2}\nabla^a\chi^b-\frac{d\epsilon^2}{4}\chi^a\chi^b\right)\nonumber\\
&+&d\epsilon^2\left(\frac{n-1}{\sigma^2}-\frac{1}{2\epsilon}\nabla_a\chi^a-\frac{d}{4}\chi_a\chi^a\right)^2\Bigg)\nonumber\\
gb_3&=&\epsilon^2\sigma^2\left(\frac{\BR}{\epsilon^2}-\frac{d}{\epsilon}\nabla_a\chi^a-\frac{d(d+1)}{4}\chi_a\chi^a+\frac{d(d-1)}{\sigma^2}\right)^2\nonumber\\
t_4&=& -2\epsilon^2\left(\frac{1}{\epsilon^2}\BR-\frac{d}{\epsilon}\nabla_a\chi^a-\frac{d(d+1)}{4}\chi_a\chi^a+\frac{d(d-1)}{\sigma^2}\right)\left(d-1-\frac{1}{2\epsilon}\nabla_a\chi^a\sigma^2-\frac{d}{4}\chi_a\chi^a\sigma^2\right)\nonumber\\
t_5&=& 4\sigma^2\epsilon^2\left(\frac{d-1}{\sigma^2}-\frac{1}{2\epsilon}\nabla_a\chi^a-\frac{d}{4}\chi_a\chi^a\right)^2\nonumber\\
t_6&=&-\frac{4\sigma^2}{\epsilon^2}\left(\BR_{ab}-\frac{d\epsilon}{2}\nabla_a\chi_b-\frac{d\epsilon^2}{4}\chi_a\chi_b\right)\left(\frac{\epsilon}{2}\nabla^a\chi^b+\frac{\epsilon^2}{4}\chi^a\chi^b\right)\nonumber\\
&-&4\epsilon^2(d-1)\left(\frac{d-1}{\sigma^2}-\frac{1}{2\epsilon}\nabla_a\chi^a-\frac{d}{4}\chi_a\chi^a\right)\left(\frac{1}{4}\chi_a\chi^a\sigma^2-1\right)\nonumber\\
t_7&=&-4\sigma^2\left(\frac{1}{2}\nabla_a\chi_b+\frac{\epsilon}{4}\chi_a\chi_b\right)\left(\frac{1}{2}\nabla^a\chi^b+\frac{\epsilon}{4}\chi^a\chi^b\right)-\frac{4(d-1)}{\sigma^2}\left(\frac{\sigma^2}{4}\chi_a\chi^a-1\right)^2
\end{eqnarray}
\section{Choice of patch coordinate}

Here we will specify a particular choice of scaled coordinates `$y^a$' \cite{Bhattacharyya:2015dva, Bhattacharyya:2015fdk}, that we will use for our computations. To start with, we note that there are three distinct directions in the geometry under consideration, namely, $1)$ the unit spacelike normal  $n_\mu$ to the $\psi=1$ surface, $2)$ the unit time-like one-form `velocity' $u_\mu$ and $3)$ the  distinct radial direction normal to the isometry directions i.e. $dS$. We will use a particular set of linear combination of these three directions as three of our basis vectors for our coordinate system namely,  $n, O=n-u$ and $Z=dS-(n.dS) n$. In addition we will use an arbitrary set of orthonormal coordinates $Y^i$ for the rest of the $p$ basis vectors. These $Y^i$ coordinate directions are also orthogonal to the hypersurface defined by the $n, O$ and the $Z$ directions. There is a lot of ambiguity in choosing the $Y^i$ coordinates but our final result will be written in a manner independent of the details of this choice. 
The scaled coordinates $y^a$ used by us is given by
\begin{eqnarray}
&&R=(D-3)(\psi-1)\nonumber\\
&&V=(D-3) \left(x^\mu-x_0^\mu\right)O_\mu\nonumber\\
&&z=(D-3)\left(x^\mu-x_0^\mu\right)Z_\mu\nonumber\\
&&y^i=(D-3)\left(x^\mu-x_0^\mu\right)Y^i_\mu
\end{eqnarray}
In these scaled coordinates ansatz metric \eqref{ansatz_metric} in the effective $p$ dimensional spacetime is given by 
\begin{eqnarray}\label{ansatzincoord}
ds^2&=&2\frac{S_0}{n_s}dVdR-\left(1-e^{-R}\right)\left(1-\beta\frac{n_s^2}{S_0^2}e^{-R}\right)dV^2+\frac{dz^2}{1-n_s^2}+dy_idy^i+\mathcal{O}(1/D)\nonumber\\
e^{\phi_0}&=&S_0^2++\mathcal{O}(1/D)
\end{eqnarray}
Our primary task will be to add corrections to the leading order ansatz metric \eqref{leadingansatz} of the form
\begin{eqnarray}\label{metric_expansion}
h_{\mu\nu}&=&\sum_{k=0}^\infty\left(\frac{1}{D-3}\right)^k\left(h_{\mu\nu}^{(k)}+\beta h_{\mu\nu,\beta}^{(k)}\right)\nonumber\\
\phi&=&\sum_{k=0}^\infty\left(\frac{1}{D-3}\right)^k\left(\phi^{(k)}+\beta\phi^{(k)}_\beta\right)
\end{eqnarray}
so that the full metric \eqref{metric_expansion} satisfies the EGB equations to desired order in $\frac{1}{D-3}$. Note that  $h_{\mu\nu}^{(0)}+h_{\mu\nu,\beta}^{(0)}$ is the leading order ansatz metric given in \eqref{ansatz_metric}. 
\section{Solving for the metric corrections}
%\subsection{Simplification of EGB equations in the large $D$ limit}
The functions $\psi$ and $u$ present in the ansatz metric have order $\mathcal{O}(1)$ derivatives in the global coordinates, hence these are almost constants with respect to the patch coordinates $y^a$ in the membrane region. But the metric has order $D$ derivatives along the $d\psi$ direction. Hence, to the leading non-trivial order in $1/D$, when the EGB equations act on the ansatz metric, they reduce to functions of the coordinate $R$ only. This can easily be seen from the leading order in $1/D$ piece of the ansatz metric written in \eqref{ansatzincoord}. The coefficients of the various terms in these functions of $R$ are composed of Taylor expansion coefficients of $\psi$ and $u$ about the particular point under consideration. So, we see that the corrections that we need to add to the ansatz metric need only be functions of $R$. Hence, the EGB equations acting on the ansatz metric plus the corrections become ordinary differential equations (ODE) in $R$ for the corrections with inhomogeneous sources formed with taylor expansion coefficients of the shape and velocity functions. The schematic form of these equations look like

%Since, the functions $\psi$ and $u$ have order $\mathcal{O}(1)$ derivatives globally, to leading order in $1/D$ these can be treated as constants in the zoomed in membrane region. We also know that the But it is obvious from \eqref{ansatzincoord} that the metric has non-trivial derivatives along $R$ as can be seen from the exponential fall off to flat spacetime in \eqref{ansatzincoord}. Hence, when the ansatz metric is fed as an input to the EGB equations and it is evaluated at the first subleading order in $1/D$ in the membrane region, we are left with an expression which is a function of $R$ only.  The coefficients of various terms in this function are composed of taylor expanstion coefficients of $\psi$ and $u$ about a particular point under consideration. This happens because in the above mentioned choice of patch coordinates, the subleading in $1/D$ terms in the EGB equation can only come about when the transverse derivatives act on the shape and velocity function in the starting ansatz metric. These transverse derivatives then manifest themselves as the taylor expansion coefficients of the shape and velocity function about the point of consideration. 

%The corrections that we need to add needs to only be function of the $R$ and hence the EGB equations evaluated on the full metric becomes an ordinary differential equation (ODE) in $R$ for the corrections with some inhomogeneous sources formed with taylor expansion coefficients of the shape and velocity functions. 

\begin{equation}
\left(\mathcal{D}_0+\beta \mathcal{D}_1\right)(h^{(1)}(R)+\beta h^{(1)}_\beta(R))=\mathcal{S}(R)
\end{equation}
where $\mathcal{D}_0$ and  $\mathcal{D}_1$ are ordinary differential operators in $R$. This simplification would not have been possible in finite $D$ as there would not have been a parametric seperation of scales between the derivatives along $d\psi$ and other directions.

\subsection{Tensor Structure of the EGB equations}
For the purpose of convenience of computation we will classify the equations according to their index structure w.r.t the directions orthogonal to the $u$, $n$ and $dS$ directions. We will classify the equations as scalars vectors and tensors depending upon the number of indices in the equations along the $y^i$ directions in the effective spacetime. Since the differential operators are linear, equations of a particular tensor structure can only contain metric corrections of that tensor structure. Similarly the source to the homogeneous parts of the differential equations must also transform under the same tensor structure. This simplifies the computation process as one can solve each sector independently, as there is no mixing between different tensor sectors. 

Here we will list out all independent objects that can appear in the inhomogeneous part of the  ODEs at the leading non-trivial order in $1/D$. We will list out the independent  objects in each tensor sector after imposing the following constraints on the shape and velocity functions.
$$n\cdot n=1,\quad n\cdot u=0 \quad \text{and} \quad u\cdot u=-1.$$
In addition we also take into account the auxiliary conditions on the shape and velocity data, namely
$$n.\nabla n=0, \quad \text \quad n.\nabla u=0$$
Our choice are the same as those of \cite{Bhattacharyya:2015dva, Bhattacharyya:2015fdk} and are given by
%We will find it convenient to categorize the equations according totheir transformation properties under the $SO(p)$ invariant $y^i$ directions of the effective background metric. Since, the homogenous part of the EGB equations at first subleading order in $1/D$ turn out to be  ODE's for the metric corrections mentioned in \eqref{explicit_correction_metric} they can be classified as scalars, vectors or tensors depending on the number of components (zero, one or two) along the $y^i$ directions. This is the underlying reason for the nomenclature used for the corrections in \eqref{explicit_correction_metric} (`S' for scalars, `V' for vectors and `T' for tensors). The advantage of this classification is that now the sources to the homogeneous parts of the equations will also transform under the corresponding tensor structure only. Hence, if we also classify the sources that appear in these equations according to the tensor structure, we will find the procedure to solve the equations much simpler. This classsification of sources has already been spelled out in great detail in  \cite{Bhattacharyya:2015dva, Bhattacharyya:2015fdk}. Below we list out the independent data in each sector that appears in the expressions for sources to the ODE's. These independent data of the membrane have been drawn up after taking into account the constraints on the data namely $$n.n=1, \quad u.u=-1\quad \text{and}\quad n.u=0$$ and the auxilliary conditions 
\subsubsection{scalar sector}\label{independent_data}
$s_1=u.K.u, \quad s_2=u.K.z,\quad s_3=z.K.z,\quad s_4=P^{\mu\nu}K_{\mu\nu}\quad \text{and}\quad  s_5=P^{\mu\nu}\nabla_\mu u_\nu$
\subsubsection{vector sector}
$v_1^\mu=u^\alpha K_{\alpha\beta}P^{\beta\mu},\quad v_2^\mu=z^\alpha K_{\alpha\beta}P^{\beta\mu},\quad v_3^\mu=u^\alpha\partial_\alpha u_\beta P^{\beta\mu},\quad  v_4^\mu=z^\alpha \partial_\alpha u_\beta P^{\beta\mu}$
\subsubsection{tensor sector}
$t_1^{\mu\nu}=P^{\mu\alpha}P^{\nu\beta}\left(K_{\alpha\beta}-\frac{P_{\alpha\beta}}{p}P^{\theta\phi}K_{\theta\phi}\right),\quad \text{and}\quad t_2^{\mu\nu}=P^{\mu\alpha}P^{\nu\beta}\left(\nabla_{(\alpha}u_{\beta)}-\frac{P_{\alpha\beta}}{p}P^{\theta\phi}\nabla_{\theta}u_\phi\right)$
\newline\newline$P^{\mu\nu}$ is a projector orthogonal to the three distinct directions in the effective $p+3$ dimensional geometry, namely $u,n$ and $dS$ and $K_{\mu\nu}$ is the extrinsic curvature of the constant $\psi$ slice corresponding to the point of interest in the membrane region. 
\subsection{Gauge choice}
We have to fix the gauge for the metric corrections to take care of the unphysical diffeomorphism degrees of freedom of the EGB equations. We will use the gauge choice of  \cite{Bhattacharyya:2015fdk, Dandekar:2016fvw} given by $$h_{MN}O^M=0.$$ The most general correction to the metric at leading order in the above mentioned coordinates system can be written as
\begin{eqnarray}\label{explicit_correction_metric}
h_{ab}^{(1)}+\beta h_{ab,\beta}^{(1)}&=&(S_{VV}(R)+\beta S_{VV,\beta}(R))dV^2+2(S_{Vz}(R)+\beta S_{Vz,\beta}(R))dVdz\nonumber\\
&+&(S_{zz}(R)+\beta S_{zz,\beta}(R))dz^2+(S_{tr}(R)+\beta S_{tr,\beta}(R))dy^idy^i\nonumber\\
&+&2(V_{Vi}(R)+\beta V_{Vi,\beta}(R))dVdy^i+2(V_{zi}(R)+\beta V_{zi,\beta}(R))dzdy^i\nonumber\\
&+&(T_{ij}(R)+\beta T_{ij,\beta})dy^idy^j.
\end{eqnarray}
Here $T_{ij}$ is traceless in the sense that $T_{ij}\delta^{ij}=0$. The letters denoting the metric corrections: $S,V$ and $T$ indicate the tensor sector to which they belong. 
\subsection{Boundary conditions and definition of velocity and shape of the membrane}
In our coordinate system the starting ansatz metric is regular everywhere except for the singularity inside the horizon. We would like to preserve this property and hence, we require that the metric corrections that we find are also regular everywhere in spacetime away from the above mentioned singularity. For regions far outside the membrane region the spacetime is flat and hence it is always regular. Since regions inside the horizon are causally disconnected from everything outside we will not be worried about the behaviour of metric corrections far inside. Hence, we need to make sure that the solutions are regular everywhere in the membrane region. 

Secondly, we will only be interested in those solutions which approach the flat space time outside the membrane region as the $d\psi$ direction is the fast direction for the metric. This can be implemented by making sure that the metric and scalar field corrections vanish as the coordinate $R$ approaches infinity.

We find that even after using the above boundary conditions we are still left with some  unfixed degrees of freedom in the solution. The ambiguity is worth one scalar plus one vector degree of freedom. The source of this ambiguity is the fact that the leading order ansatz solution is invariant if we redefine the shape function by $\psi\rightarrow\psi+\frac{1}{D}\delta\psi$ and the velocity function by $u\rightarrow u+\frac{1}{D}\delta u$. But these redefinitions add $\mathcal{O}(1/D)$ piece to the ansatz metric and hence they modify the corrections to the metric at $\mathcal{O}(1/D)$. Hence, to unambiguously specify the corrections we must give a precise definition of the shape and the velocity field. This is similar to choice of frame in fluid-gravity correspondence. 

We adopt the geometric definitions of the shape and velocity fields, proposed in \cite{Dandekar:2016fvw}. We demand that the position of the event horizon coincides with the membrane surface $\psi=1$ \footnote{The surface $\psi=1$ is the event horizon for the starting ansatz metric to leading order in $1/D$. Hence, we expect that the horizon of the metric with the subleading corrections to be in the vicinity of width $1/D$ about this surface. Since the original surface was a null hypersurface, we expect that the horizon of the corrected metric which is $1/D$ shifted from its zeroth order position to still be a null surface. We have not exactly found the horizon surface by extrapolating null rays from future infinity, but we expect that this procedure will also lead to the same surface. We thank Benjamin Withers for discussions on this matter}. Hence, at $\psi=1$ the normal to the surface  must be a null vector w.r.t. the full metric. If $G_{MN}$ is the full metric of the spacetime at a desired order in $1/D$ then the condition that the $\psi=1$ surface is event horizon is given by $$G^{MN}\partial_M\psi\partial_N\psi|_{\psi=1}=0$$
Similarly, the  velocity field $u$ is the generator of the event horizon for the starting ansatz metric and we demand that this property holds for all orders in $1/D$, i.e 
$$u^A=G^{AB}n_B.$$ The above two conditions can easily be shown (see \cite{Dandekar:2016fvw}) to result in the following conditions on the metric corrections. 
\begin{eqnarray}\label{boundary condition}
S_{VV}(0)+S_{VV,\beta}(0)=0\nonumber\\
S_{Vz}(0)+S_{Vz,\beta}(0)=0\nonumber\\
V_{Vi}(0)+V_{Vi,\beta}(0)=0
\end{eqnarray}
Once the regularity and boundary conditions are imposed, the final solution is unambiguously expressed in terms of the shape and velocity function of the membrane\footnote{We also need to define the auxiliary conditions on the shape and velocity field which in our case is given by \eqref{auxilliary_condn_psi} }. 
\subsection{The Constraint EGB Equations and the Equations of motion of the Membrane}
As we mentioned earlier, the EGB equations are invariant under diffeomorphisms. To fix this ambiguity we chose a particular gauge for the metric corrections which reduce the number of independent components in the correction metric from $\frac{(p+3)(p+4)}{2}$ to $\frac{(p+2)(p+3)}{2}$ (in addition we have one scalar correction), though the number of components of the EGB equations remain $\frac{(p+3)(p+4)}{2}$ (plus one scalar equation). So, naively, it looks like that we have a over-determined system at hand. A Similar situation arises in two derivative Einstein gravity and there the solution to this apparent problem is well known. There the solution lies in the fact that not all components of the Einstein's equations track the  `evolution' of a metric from a set of arbitrary initial data on a hypersurface. There are exactly $p+3$ components of the equations which constrain the `initial' data on the hypersurface itself. These equations put constraints on the  possible initial data which one can use. Once a consistent set of initial data has been defined, rest of the $\frac{(p+2)(p+3)}{2}$ equations dynamically evolve this `initial' data to give the metric correction at any point in spacetime.  In addition once the constraint equations are solved on any given hypersurface (normal to the direction of evolution) and the dynamical equations are solved everywhere, then the constraint equations are automatically satisfied on any hypersurface orthogonal to the direction of evolution.  It was shown in \cite{ChoquetBruhat:1988dw} that a similar structure of constraint and dynamical equations exist for EGB gravity also.

Since we will be interested in studying `evolution' of the metric components using the EGB equations along the normal to the membrane direction, the constraint equations in this case are
\begin{equation}
C_{\mu}=E_{\mu\nu}G^{\nu\beta}n_\beta
\end{equation}
where $n^\mu$ is the normal to the $\psi=1$ hypersurface and $G_{\mu\nu}$ is the full metric at the desired order in $1/D$. $E_{\mu\nu}$ are the EGB equations. In the classification of EGB equations according to their tensor structure mentioned above, there are two distinct type of constraint equation: scalar and vector. We will explicitly write down the constraint equations in both these sectors below and also demonstrate following arguments similar to \cite{Dandekar:2016fvw} that these equations along with the imposition of the condition that the metric corrections are regular everywhere, gives rise to a set of constraints on the shape and velocity data on the membrane. We will call these the membrane equations of motion.

%In our classification of quantities according to their tensor structure under the $SO(p)$ transformation properties in the directions orthogonal to $n, u$ and $dS$ directions we choose the following independent combinations of scalar and vector constraint equations following \cite{Dandekar:2016fvw}. 

\subsubsection{Scalar constraint equations}
There are three scalar constraint equations formed by taking the components of the constraint equations $C_\mu$ along $u, dz$ and $O$ directions. We explicitly write down the homogeneous parts of the constraint equations along $u$ and $dz$ directions below as these will determine the constraints on the scalar data coming from the shape and velocity of the membrane. The constraint equation along $O$ direction gives no additional constraint on the data. The relevant scalar constraint equations are given by
\begin{eqnarray}\label{scacons1}
C_\mu u^{\mu}&=&-\frac{n_S(1-n_S^2)}{2S_0^2}\left(e^{-R}S_{Vz}-(1-e^{-R})\frac{d S_{Vz}}{dR}\right)\nonumber\\
&&-\beta \frac{n_S(1-n_S^2)}{2S_0^4}\left(n_S^2e^{-R}S_{Vz}+e^{-R}S_0^2S_{Vz\beta}-(1-e^{-R})\left(n_S^2e^{-R}\frac{d S_{Vz}}{dR}+S_0^2\frac{d S_{Vz\beta}}{dR}\right)\right)\nonumber\\
&&+S_{const}^u(R)=0
\end{eqnarray}
\begin{eqnarray}\label{scacons2}
C_z&=& \frac{n_S(1-n_S^2)}{2S_0^2}(1-e^{-R})\frac{d S_{zz}}{dR}\nonumber\\
&&+\beta\frac{n_S(1-n_S^2)}{2S_0^4}(1-e^{-R})\left(-n_S^2e^{-R}\frac{d S_{zz}}{dR}+S_0^2\frac{d S_{zz\beta}}{dR}\right)+S_{const}^z(R)=0~~~~~
\end{eqnarray}
$S^u_{const}(R)$ and $S_{const}^z(R)$ schematically denote the sources for these ODEs which are formed from the independent scalar data listed in \eqref{independent_data}
\subsubsection{Vector constraint equations}
The expression for the vector constraint equations is given by
\begin{eqnarray}
C_\mu^VP^\mu_i&=&\frac{n_S(1-n_S^2)}{2S_0^2}(1-n_S^2)(1-e^{-R})\frac{d V_{zi}}{dR}\nonumber\\
&&+\beta \frac{n_S(1-n_S^2)}{2S_0^4}(1-e^{-R})\left(-n_S^2e^{-R}\frac{d V_{zi}}{dR}+S_0^2\frac{d V_{zi\beta}}{dR}\right)+S_{const,i}^V(R)=0\nonumber\\
\end{eqnarray}
$S_{const,i}^V$ are the sources for the vector constraint equations formed using the vector data listed in \eqref{independent_data}

Notice that in both the scalar and vector constraint equations the order of the equations is one. For the rest of the dynamical equations the order of the ODE's will be 2. This is in agreement with the discussion above  about the constraint equations not being dynamical in nature. To completely determine the metric using the EGB equations we need to specify the metric components and their normal derivative as the initial data on a given hypersurface. And the constraint equation being only first order imposes restrictions on this initial data.  
\subsubsection{The Membrane equations of motion}
Since the constraint equations can be solved on any constant $R$ slice, we will chose the surface $R=0$ for convenience. The regularity of the metric corrections everywhere in the membrane region makes the homogeneous parts of the equations go to zero at $R=0$ and we arrive at the following two constraints on the scalar data
\begin{eqnarray}
&&S^u_{const}(0)=0\nonumber\\
&&S^z_{const}(0)=0\nonumber\\
\end{eqnarray}
The explicit form of the constraints on the shape and the velocity data is given by
\begin{eqnarray}\label{scalar_mem_eq_S}
&&u\cdot dS=-\frac{S_0}{D}\partial_\mu u^\mu\nonumber\\
&&s_3\left(1-\alpha \frac{n_s^2}{S_0^2}\right)-2 s_2+s_1\left(1+\alpha \frac{n_s^2}{S_0^2}\right)-\frac{1-n_s^2}{n_s S_0}\left(1-\alpha \frac{n_s^2}{S_0^2}\right)=0
\end{eqnarray}
We call the above two equations the scalar membrane equations of motion. Similar arguments gives rise to the following constraint on the vector data
\begin{equation}
S_{const,i}^V=0
\end{equation}
Written explicitely this gives
\begin{equation}\label{vec_mem_eq_S}
\left(v_4^\mu-v_3^\mu\left(1+\alpha \frac{(n.dS)^2}{S_0^2}\right)-v_2^\mu\left(1-\alpha \frac{(n.dS)^2}{S_0^2}\right)+v_1^\mu\right)P^\mu_i=0
\end{equation}
We call this the vector membrane equation of motion. 
%We expect that the solution that we find are regular everywhere in the membrane region and hence in particular also behave smoothly at $R=0$. Assuming this we see that the homogeneous parts of the above equations vanish at $R=0$. Since, the constraint equations has to be satisfied at any $R=const$ slice hence, demanding that they be satisfied at $R=0$ gives the following constraint on the sources mentioned above 
%\begin{eqnarray}
%&&S_{const}(0)^u=0\nonumber\\
%&&S_{const}(0)^z=0\nonumber\\
%&&S_{const,i}^V=0
%\end{eqnarray}
%The sources are formed from the shape and velocity data of the membrane and hence, the above constraint on the sources gives rise to constraint on the shape and velocity field on the membrane. We call these the `membrane equation of motion'. We will list them below
%\subsubsection{scalar equation}
%\begin{eqnarray}
%&&u_S=-\frac{S_0}{D}\partial_\mu u^\mu\nonumber\\
%&&s_3\left(1-\alpha \frac{n_s^2}{S_0^2}\right)-2 s_2+s_1\left(1+\alpha \frac{n_s^2}{S_0^2}\right)-\frac{1-n_s^2}{n_s S_0}\left(1-\alpha \frac{n_s^2}{S_0^2}\right)=0
%\end{eqnarray}
%\subsubsection{vector equation}
%\begin{equation}
%\left(z.\partial u_j-u.\partial u_j\left(1+\alpha \frac{n_s^2}{S_0^2}\right)-z^mK_{mj}\left(1-\alpha \frac{n_s^2}{S_0^2}\right)+u^mK_{mj}\right)P^j_i=0
%\end{equation}
\subsection{The dynamical equations and the leading order metric correction}
In the above section we have used the constraint EGB equations to impose the condition of regularity on the metric corrections on the $R=0$ slice and we found that to be able to impose this condition, the shape and velocity data on the membrane must satisfy certain constraints given in \eqref{scalar_mem_eq_S} and \eqref{vec_mem_eq_S}. Once the constraint equations are taken care of we can now proceed to solve for the metric corrections using the dynamical equations. We will not explicitly write down the dynamical EGB equations here and just write down the final solutions after all the boundary conditions have been imposed. 

%From the structure of the EGB equations we know that the constraint equations need to be satisfied on only one of the $R=constant $ slices. We found that the $R=0$ slice was particularly useful to us and using that we were able to constraint the data on the dual membrane in the form of the `membrane equation of motion'. Having done that, we have made sure that there are no more obstructions to finding unique solutions to the EGB equations once we keep ourselves confined to the boundary data that is allowed by the constraint equation. Here we will list out the various components of the leading order in $1/D$ corrections to the metric which satisfy the boundary conditions \eqref{boundary condition}
\begin{eqnarray}
S_{VV}&=&-e^{-R}R^2+\frac{n_S}{S}s_1\left(e^{-R}~R+\frac{e^{-R}~R^2}{2}\right)-\frac{ S}{n_S^2}s_3 R^2e^{-R}\nonumber\\
&&+\beta\bigg(\frac{n_S^2}{S^2}\left(e^{-2R}(1+2R(R-1))-e^{-R}(R-1)^2\right)-2s_2 S R(R-1)(e^R-1)e^{-2R}\nonumber\\
&&+\frac{n_S}{2S}s_1\left(-2e^{-2R}(R^2+2)+e^{-R}(4+2R+3R^2)\right)\bigg)\nonumber\\
S_{Vz}&=&\frac{R~e^{-R}}{(1-n_S^2)}\left(-\frac{s_2}{n_S}+s_1\right)+\beta \frac{n_S^2}{S^2}\frac{R(3e^{R}-1)e^{-2R}R}{(1-n_S^2)^2}\left(-\frac{s_2}{n_S}+s_1\right)\nonumber\\
S_{zz}&=& \beta \frac{2 e^{-R}}{(1-n_S^2)^2}\frac{n_S^2}{S^2}\left(-\frac{s_2}{n_S}+s_1\right)\nonumber\\
S_{Tr}&=& \beta ~e^{-R}\left(2\frac{n_S^2}{S^2}+\frac{n_S}{S}(s_4+s_5)\right)\nonumber\\
\delta \phi^{(2)}&=&-2S_{Tr}-(1-n_S^2)S^2S_{zz}\nonumber
\end{eqnarray}
\begin{eqnarray}
V_{z i}&=&\beta \frac{e^{-R}}{1-n_S^2}\frac{n_S}{S}\left(v_{3\alpha}n_S-v_{2\alpha}\right)P^{\alpha}_i\nonumber\\
V_{Vi}&=&\frac{R~e^{-R}S}{n_S}\left(v_{1\alpha}-v_{3\alpha}\right)P^\alpha_{i}\nonumber\\
&&+\beta R~e^{-2R}\left(v_{2\beta} e^R+\frac{n_S}{S}v_{1\beta}(e^R-1)+\frac{n_S}{S}(1-2e^R)v_{3\beta}\right)P^{\beta}_{\mu}\nonumber\\
T_{\mu\nu}&=&2\beta \frac{n_S}{S}e^{-R}\left(t_{1\mu\nu}-t_{2\mu\nu}\right)\nonumber\\
\text{where}&&\quad P_{\mu\nu}=g_{\mu\nu}+u_\mu u_\nu-n_\mu n_\nu-\frac{z_{\mu}z_\nu}{1-n_S^2}\quad \text{and}\quad n_S=n.dS
\end{eqnarray}
\section{Geometric form of the metric and the membrane equation of motion}
We have written down the metric corrections and the membrane equations in the effective $p+3$ dimensional spacetime. The computation has been done using computer algebra. Following \cite{Bhattacharyya:2015fdk, Bhattacharyya:2015dva, Dandekar:2016fvw} we do the above computation for different values of $p=2,3$. The results can be recast in terms of shape and velocity data of the membrane in full spacetime (obviously preserving the huge isometry). The procedure for this has been explained in great details in \cite{Bhattacharyya:2015fdk} and we will not elaborate on that here. This way of writing the answer in terms of the shape ane velocity data of the full $D$ dimensional spacetime was called as the geometric form of the solution. We write down below our answer in the geometrical form and it turns out to be independent of $p$. 
\subsection{Metric correction in geometric form}
The most general form of the first order corrected metric in the geometric form (with our choice of gauge) is given by
\begin{eqnarray}\label{geom_metric}
ds^2=ds_{0}^2+\frac{1}{D}\left(H_s(O_A dx^A)^2+H_A^{(V)}O_B dx^A dx^B+H_{AB}^{(T)}dx^A dx^B+\frac{1}{D}H^{Tr}P_{AB}dx^A dx^B\right)\nonumber\\
\end{eqnarray}
The metric corrections int he effective spacetime written in terms of the full spacetime metric mentioned above is given by
\begin{eqnarray}
H_s=&=&-e^{-R}R^2+\frac{\mathcal{K}}{D-3}u.K.u\left(e^{-R}~R+\frac{e^{-R}~R^2}{2}\right)-\frac{ D-3}{\mathcal{K}}\frac{u.\nabla\mathcal{K}}{\mathcal{K}} R^2e^{-R}\nonumber\\
&&+\beta\bigg(\frac{\mathcal{K}^2}{(D-3)^2}\left(e^{-2R}(1+2R(R-1))-e^{-R}(R-1)^2\right)\nonumber\\&&-2\frac{u.\nabla\mathcal{K}}{D-3} R(R-1)(e^R-1)e^{-2R}\nonumber\\
&&+\frac{\mathcal{K}}{2(D-3)}u.K.u\left(-2e^{-2R}(R^2+2)+e^{-R}(4+2R+3R^2)\right)\bigg)\nonumber
\end{eqnarray}
\begin{eqnarray}\label{geom_metric_exp}
H^{(V)}_M&=&\frac{R~e^{-R}(D-3)}{\mathcal{K}}\left(u^B K_{BA}-u.\nabla u_{A}\right)P^A_{M}\nonumber\\
&&+\beta R~e^{-2R}\left(\frac{\nabla_B\mathcal{K}}{D-3} e^R+\frac{\mathcal{K}}{D-3}u^{A}K_{AB}(e^R-1)+\frac{\mathcal{K}}{D-3}(1-2e^R)u.\nabla u_B\right)p^{B}_{M}\nonumber\\
H^{(T)}_{MN}&=&2\alpha\frac{\mathcal{K}}{D-3}\mathcal{T}_{MN}\nonumber\\
\text{where}\quad\mathcal{T}_{MN}&=&P_M^A P_{N}^{B}\left(K_{AB}-U_{AB}\right)-\frac{P_{MN}}{D-3}P^{TF}\left(K_{TF}-U_{TF}\right)\nonumber\\
\text{and}\quad U_{AB}&=&\frac{\nabla_M u_N+\nabla_N u_M}{2}\nonumber\\
H^{Tr}&=&\mathcal{O}\left(\frac{1}{D^3}\right)\nonumber\\
\text{where}\quad &&R=D(\psi-1)
\end{eqnarray}
$K_{MN}$ is the extrinsic curvature of the membrane surface $\psi=1$ embedded in the full flat spacetime. 
\subsection{Membrane equations in the geometric form}
Similarly, the membrane equations of motion of \eqref{scalar_mem_eq_S} and \eqref{vec_mem_eq_S} can be recast in a geometrical form. In this language there is a single scalar equation obtained from the first line of \eqref{scalar_mem_eq_S} given by
\begin{equation}\label{geomtric_scal_eqn}
\nabla.u=0
\end{equation}
and a single vector equation obtained from \eqref{vec_mem_eq_S} given by
\begin{equation}\label{geometric_vect_eqn}
\mathcal{P}_A^M\left(\frac{\nabla^2 u_{M}}{\mathcal{K}}+u^{B}K_{BM}-u.\nabla u_M\left(1+\beta \frac{\mathcal{K}^2}{(D-3)^2}\right)-\frac{\nabla_M\mathcal{K}}{\mathcal{K}}\left(1-\beta\frac{\mathcal{K}^2}{(D-3)^2}\right)\right)=\mathcal{O}\left(\frac{1}{D}\right)
\end{equation}
The second of \eqref{scalar_mem_eq_S} is simply the divergence of the above vector equation ( as was the case in \cite{Bhattacharyya:2015fdk}) and is given by
\begin{equation}\label{divofeq}
\frac{\nabla^2\mathcal{K}}{\mathcal{K}^2}\left(1-\beta\frac{\mathcal{K}^2}{(D-3)^2}\right)+u.K.u \left(1+\beta \frac{\mathcal{K}^2}{(D-3)^2}\right)-2u.\nabla \mathcal{K}=\mathcal{O}\left(\frac{1}{D}\right)
\end{equation}
This last equation contains no new information. 
\section{A world volume stress tensor}
By comparing the equations of motion of the membranes dual to black holes in EGB gravity \eqref{geometric_vect_eqn} and \eqref{divofeq} with the corresponding quantities of \cite{Bhattacharyya:2015fdk} we see that they map to each other under the substitution 
$$Q\rightarrow \sqrt{\beta}\frac{\mathcal{K}}{D}.$$
We have already noticed that the starting ansatz metric for both EGB gravity and Einstein-Maxwel theory are the same under the above substitution. Taking cue from this we propose that the stress tensor for the membrane dual to EGB black holes at leading order in $1/D$ can be obtained from the stress tensor for membranes dual to charged black holes derived in \cite{Bhattacharyya:2016nhn}, upon using the above mentioned substitution. The proposed leading order stress tensor for EGB gravity is then given by\footnote{We have neglected terms in the stress tensor whose divergence gives $\mathcal{O}(1)$ quantities. This is so because we will be proving that the divergence of stress tensor which naively should have been atleast $\mathcal{O}(D^2)$ is actually only $\mathcal{O}(D)$ if equation of motion of the membrane is satisfied. We will not be interested in finding the explicit form of the $\mathcal{O}(1)$ quantity that will be left out in the conservation equation.}
\begin{eqnarray}\label{mem_stress_tensor}
8\pi T_{MN}&=&\frac{\mathcal{K}}{2}\Bigg(1+\frac{\beta\mathcal{K}^2}{D^2}\Bigg)u_M u_N+\Bigg(1-\frac{\beta\mathcal{K}^2}{D^2}\Bigg)\frac{K_{MN}}{2}-\frac{\nabla_M u_N+\nabla_N u_M}{2}-\left(u_M V_N+u_N V_M\right)\nonumber\\
\text{where}&&\nonumber\\
V_M&=& -\frac{1}{2}\left(1-\frac{\beta\mathcal{K}^2}{D^2}\right)\frac{\nabla_M\mathcal{K}}{\mathcal{K}}+\frac{\beta\mathcal{K}^2}{D^2}u^A K_{AM}-\frac{\beta\mathcal{K}^2}{2D^2}u\cdot \nabla u_M+\left(1+\frac{\beta\mathcal{K}^2}{D^2}\right)\frac{\nabla^2u_M}{\mathcal{K}}\nonumber\\
\end{eqnarray}
This stress tensor is the world-volume stress tensor of the membrane and hence we will only be interested in the equation of conservation of this stress tensor in the world volume of the membrane moving in flat spacetime. The conservation equation has one free index and this can either be along the velocity field $u$ (which is the only distinct direction in the world-volume), given by
\begin{equation}
E=8\pi \nabla^M T_{MN}u^N=0,
\end{equation}
or it could be in a direction orthogonal to the velocity field given by
\begin{equation}
E_N=8\pi \nabla^M T_{MA}P^{A}_N=0
\end{equation}
The explicit expression for $E$ is 
\begin{eqnarray}\label{scalar_conservation}
E&=&-\frac{\mathcal{K}}{2}\left(1+\frac{\beta\mathcal{K}^2}{D^2}\right)\nabla\cdot u-\frac{\beta\mathcal{K}^2}{2D^2}u\cdot\nabla\mathcal{K}-\frac{1}{2}\left(1-\frac{\beta\mathcal{K}^2}{D^2}\right)\frac{\nabla^2\mathcal{K}}{\mathcal{K}}\nonumber\\
&&-\frac{\beta\mathcal{K}^2}{2D^2}\nabla^M(u\cdot \nabla u_M)+\left(1+\frac{\beta\mathcal{K}^2}{D^2}\right)\frac{\nabla^M(\nabla^2 u_M)}{\mathcal{K}}
\end{eqnarray}
To obtain the above expression we have used one of the Gauss-Codazzi equations for the surface embedded in flat spacetime (we call this the equation of the first type) given by $$\nabla^M K_{MN}=\nabla_N\mathcal{K}.$$ We massage the last two terms of the equation\eqref{scalar_conservation} to turn them into more familiar objects. First of all
\begin{eqnarray}
\nabla^M(u\cdot \nabla u_M)&=&R_{BA}u^{B}u^{A}+\mathcal{O}(1)\nonumber \\
&=&\mathcal{K}u.K.u+\mathcal{O}(1)
\end{eqnarray}
The first equality is obtained by neglecting the term with derivative on $u$ as it is $\mathcal{O}(1)$. Then we commute the derivatives and use the fact that $\nabla\cdot u=\mathcal{O}(1)$ for the leading order ansatz metric. The second line is obtained by using the second kind of Gauss-Codazzi equation for a flat spacetime background given by
$$R_{ABCD}=-K_{AD} K_{BC}+K_{BC}K_{BD}$$
where $R_{ABCD}$ is the Riemann tensor of the induced metric on world-volume. Finally we neglect a term of $\mathcal{O}(1)$ to obtain the last line of the above equality. Also,  
\begin{eqnarray}
\nabla^M\nabla^2u_M&=&\nabla^A\nabla_M\nabla_A u^M\nonumber\\
&=& \nabla^A(R_{BA} u^B) +\mathcal{O}(1)\nonumber\\
&=&\mathcal{K}u\cdot\nabla \mathcal{K}
\end{eqnarray}
The first equality is obtained from the fact that the commutator of second and third derivative turns out to be zero.~The second equality is obtained because $\nabla\cdot u$ is  $\mathcal{O}(1)$ and third line is obtained by using both of the Gauss-Codazzi equations or flat background. Now substituting the above two expressions in \eqref{scalar_conservation} we obtain
\begin{equation}
E=-\frac{\mathcal{K}}{2}\left(1+\frac{\beta\mathcal{K}^2}{D^2}\right)\nabla\cdot u+u\cdot \nabla \mathcal{K}-\frac{1}{2}\frac{\nabla^2\mathcal{K}}{\mathcal{K}}\left(1-\frac{\beta\mathcal{K}^2}{D^2}\right)-\left(1+\frac{\beta\mathcal{K}^2}{D^2}\right)\frac{\mathcal{K}}{2}u.K.u
\end{equation}
If we now use the equation obtained by taking the divergence of the membrane equation of motion \eqref{divofeq} we get
\begin{equation}
E=-\frac{\mathcal{K}}{2}\left(1+\frac{\beta\mathcal{K}^2}{D^2}\right)\nabla\cdot u=0
\end{equation}
But we already know that $$\nabla\cdot u=\mathcal{O}\left(\frac{1}{D}\right)$$ and hence we have $$E=\mathcal{O}(1),$$ which is one order lower than the naive expectation. Hence, within the order of $1/D$ that we are interested in, $E=0$ provided the membrane equation to leading order is satisfied.

Let us now shift our attention to $E_M$ which is given by
\begin{eqnarray}\label{vect_conservation}
E_M=P_M^N\left(\frac{\mathcal{K}}{2}\left(1+\frac{\beta\mathcal{K}^2}{D^2}\right)u\cdot\nabla u_N+\frac{\nabla_N\mathcal{K}}{2}\left(1-\frac{\beta\mathcal{K}^2}{D^2}\right)-\frac{\nabla^2u_N}{2}-\frac{\nabla^A\nabla_N u_A}{2}\right)\nonumber\\
\end{eqnarray}
In the above expression we have already neglected terms which are $\mathcal{O}(1)$ and have also used first of the Gauss-Codazzi equations. We massage the last term using similar arguments as above to get
\begin{eqnarray}
\nabla^A\nabla_N u_M P^N_M&=&R_{BN}u^B P^N_A\nonumber\\
&=& \mathcal{K}K_{BN}u^B P^N_A.
\end{eqnarray} 
If we substitute the above expression \eqref{vect_conservation} we find that it is of the order $\mathcal{O}(1)$ since the naive order $\mathcal{O}(D)$ terms is proportional to the equation of motion of the membrane. Hence, we conclude that to leading order in $1/D$ the proposed world volume stress tensor of the membrane is conserved provided the membrane equations are satisfied to leading order in $1/D$ or vice-versa. 

\section{Quasi-Normal Modes from Linearised Membrane Equation}
In \cite{Bhattacharyya:2015dva, Dandekar:2016fvw} the spectrum of linearised fluctuations of the membrane dual to black holes in Einstein-Hilbert gravity was found to leading and first subleading order in $1/D$. It was demonstrated that the spectrum exactly reproduced the light quasi-normal mode spectrum of \cite{Emparan:2014aba} to relevant order in $1/D$. Similar analysis was undertaken in \cite{Bhattacharyya:2015fdk} and a prediction for light QNMs of static black holes in Einstein-Maxwel systems was given. Here we undertake a similar analysis and give a prediction for the light QNMs of static, spehrically symmetric black holes in EGB gravity. 
A static,  spherical membrane in spehrical polar coordinates is given by
$$r=1\quad \text{and}\quad  u=-dt$$
Let us parametrise small fluctuations about this static configurations by
\begin{equation}
r=1+\epsilon \delta r(t,\theta), \quad \text{and} \quad u=-dt+\epsilon \left(\delta u_t(t,\theta)dt+\delta u_a(t,\theta)d\theta^a\right)
\end{equation}
We will now evaluate the membrane equations to first order in $\epsilon$. The various terms in the equation evaluate to
\begin{eqnarray}
\frac{\nabla^2u_a}{\mathcal{K}}&=&\epsilon\left(-\frac{\partial_t^2\delta u_a+\nabla_{(s)}^2\delta u_a}{D}+\frac{\nabla_a\partial_t\delta r}{D}\right)\nonumber\\
\frac{\nabla_a\mathcal{K}}{\mathcal{K}}&=&\epsilon\left(\frac{\nabla_a^{(s)}\partial_t^2\delta r}{D}-\nabla_a^{(s)}\delta r -\frac{\nabla_a\nabla_{(s)}^2\delta r}{D}\right)\nonumber\\
\left(u.K\right)_a&=&-\epsilon \left(\partial_t\partial_a\delta r+\delta u_a\right)\nonumber\\
u.\nabla u_a&=&\epsilon \partial_t\delta u_a
\end{eqnarray}
A covariant derivative with a subscript or superscript $(s)$ denote the covariant derivative w.r.t. the $D-2$ dimensional unit sphere metric. Using the above expressions the linearised membrane equations become
\begin{eqnarray}\label{linememeqn}
&&\left(1+\frac{\nabla_{(s)}^2}{D}\right)\delta u_a+(1-\beta)\nabla_a^{(s)}\left(1+\frac{\nabla_{(s)}^2}{D}\right)\delta r-\partial_t\partial_a\delta r-(1+\beta)\partial_t\delta u_a=0\nonumber\\
&&~~~~~~~~~~~~\nabla_a^{(s)}\delta u^a=-D\partial_t \delta r
\end{eqnarray}
Taking a divergence of the first equation w.r.t the unit sphere metric and then using the second equation we get
\begin{equation}\label{divlinmemeqn}
\left(2+\frac{\nabla^2}{D}\right)(-D\partial_t\delta r)+(1-\beta)\nabla_{(s)}^2\left(1+\frac{\nabla^2}{D}\right)\delta r-\partial_t\nabla_{(s)}^2\delta r+(1+\beta)D\partial_t^2\delta r=0
\end{equation}
The shape fluctuation can be decomposed into a basis of spherical harmonics w.r.t the $D-2$ dimensional unit sphere in the following manner
\begin{equation}
\delta r(t,\theta)=\sum_{l=0}^{\infty}\delta r_l Y_l(\theta)e^{-iw_lt}.
\end{equation}
This when substituted in the linearised membrane equations above gives a quadratic equation for the frequencies $w_s$ for the scalar mode perturbations at a given angular momentum number which when solved gives 
\begin{equation}\label{scalar_qnm}
w_{s,l}=(1-\beta)\left(-i(l-1)\pm\sqrt{l-1}\right)
\end{equation}
The velocity fluctuation $\delta u_a$ can be decomposed into a pure vector (divergence-less) part and a part that is derivative of a scalar function w.r.t. the unit sphere, 
\begin{equation}
\delta u_a=v_a+\nabla_a\phi \quad , \text{where} \quad v_a=\sum_{l=1}^{\infty}v_{l}V_{a,l}(\theta)e^{-iw_l t}
\end{equation}
Substituting this in the equation \eqref{linememeqn} and upon using the solution of $\delta r$ from above gives the  frequency for the vector mode perturbations to be
\begin{equation}\label{vector_qnm}
w_{v,l}=-i(l-1)(1-\beta)
\end{equation}
The scalar modes described above are in agreement with the linearised fluctuations about a static spherically symmetric black hole in flat spacetime for EGB gravity obtained in equation $(4.18)$ of \cite{Chen:2017hwm} from an independent method of obtaining large $D$ effective equations. We could not find a corresponding solution for the vector modes. There are no analytic expressions of light quasinormal modes in EGB gravity for the particular scaling of the GB parameter we worked with from the gravity side and hence, we could not check our solutions against anything else. But taking cue from the Einstein-Hilbert case we propose \eqref{scalar_qnm} and \eqref{vector_qnm} to be the predictions for the light QNM spectrum of Static black holes in EGB gravity. 

For the scalar modes for $l=0$ we get $$w_s=0\quad \text{or}\quad2i (1-\beta),$$but only the $w_s=0$ is compatible with the second equation of \eqref{linememeqn}. This zero mode corresponds to the time independent scaling of the spherical membrane which is trivially a solution of membrane equation. For $l=1$ both the roots vanish and we seperately look at this mode in equation \eqref{divlinmemeqn} which becomes
\begin{eqnarray}
&&-D\partial_t\delta r+D\partial_r\delta r+(1+\beta)D\partial_t^2\delta r=0\nonumber\\
&&\implies\delta r_{l=1}=A t+B 
\end{eqnarray}
These two independent zero modes correspond to a time independent translation ( when $A=0$) and an infinitesimal boost of the black hole when ($B=0$). 

Similarly, there is a zero mode for the vector perturbations at $l=1$ which simply correspond to imparting an infinitesimal rotation to the black hole . This line of discussion remains unaffected in the presence of a GB parameter as can be seen from similar observations in \cite{Bhattacharyya:2015fdk, Dandekar:2016fvw}.  
%We must stress on the fact that this stress tensor is purely a guess based on observation of similarity with Einstein-Maxwel black holes which gives the right membrane equations. Ideally one should follow the rigorous algorithm of \cite{Bhattacharyya:2016nhn} based on matching the stress tensor with the source for gravitational radiation at asymptotic infinity. We will postpone addressing this issue to some project we plan to undertake in future. 
\section{Stationary Solutions}
In this section we will find the effective equations corresponding to membranes with a stationary configuration. These membranes will be dual to stationary black holes in large $D$. The stationary membrane configurations are the ones where the membrane has a time-like Killing direction \cite{Dandekar:2017aiv}. By this we mean that the shape of the membrane defined by the trace of the local extrinsic curvature is invariant along the time like Killing direction. If the killing vector is given by $k^M\partial_M$ say, then
$$ k^M\partial_M \mathcal{K}=0,$$
where, $k_M$ satisfies $$\nabla_M k_N+\nabla_N k_M.$$
The membrane velocity in the stationary configurations are proportional to the time like killing vector, namely,
$$u_M=\gamma k_M, \quad \text{where,}\quad \gamma =\frac{1}{\sqrt{-k^M k_M}}$$
Hence, under this configuration
\begin{equation}
t_{MN}=\frac{\nabla_M u_N+\nabla_N u_M}{2}=\frac{k_N\nabla_M \gamma+k_M \nabla_N\gamma}{2}
\end{equation}
so, that $$\nabla^M t_{MN}P^N_Q=\frac{k_N\nabla^2\gamma+\nabla^Mk_M\nabla_N\gamma}{2}P^N_Q=\mathcal{O}(D^0)$$
since, the $P^N_Q k_N=0$ (as velocity vector is proportional to killing vector) and $\nabla_Mk^M=0$ as $k^M\partial_M$ is a killing vector. 
The term $t_{MN}$ defined above is part of the membrane stress tensor given by \eqref{mem_stress_tensor}. From the above equation it is clear then that $t_{MN}$ does not contribute to the equation of motion coming from the conservation of stress tensor in the stationary configuration. 

Hence, the membrane equation in the stationary configuration to leading order in large $D$ is given by
\begin{equation}
\mathcal{P}_A^M\left(u.\nabla u_M\left(1+\beta \frac{\mathcal{K}^2}{(D-3)^2}\right)+\frac{\nabla_M\mathcal{K}}{\mathcal{K}}\left(1-\beta\frac{\mathcal{K}^2}{(D-3)^2}\right)\right)=0
\end{equation}
In the stationary configuration 
\begin{eqnarray}
u.\nabla u_M&=&\gamma^2k.\nabla k_M \quad( \because k.\nabla\gamma=0)  \nonumber\\
&=& -\frac{\gamma^2}{2}\nabla_M(k\cdot k) \quad (\because \nabla_M k_n+\nabla_N k_M=0)\nonumber\\
&=&-\frac{\nabla_M \gamma}{\gamma} \quad (\because k.k=-\frac{1}{\gamma^2})
\end{eqnarray}

Hence, the membrane equation in stationary configurations become
\begin{equation}\label{stat_eqn}
\mathcal{P}_A^M\left(-\frac{\nabla_M\gamma}{\gamma}\left(1+\beta \frac{\mathcal{K}^2}{(D-3)^2}\right)+\frac{\nabla_M\mathcal{K}}{\mathcal{K}}\left(1-\beta\frac{\mathcal{K}^2}{(D-3)^2}\right)\right)=0
\end{equation}
The above equation can be rewritten as
\begin{eqnarray}
&&\mathcal{P}_A^M\left(-\frac{\nabla_M\gamma}{\gamma}+\frac{\nabla_M\mathcal{K}}{\mathcal{K}}\frac{\left(1-\beta\frac{\mathcal{K}^2}{(D-3)^2}\right)}{\left(1+\beta \frac{\mathcal{K}^2}{(D-3)^2}\right)}\right)=0\nonumber\\
\implies &&\mathcal{P}_A^M\nabla_M\log\left(\gamma\left(\frac{1}{\mathcal{K}}+\frac{\beta \mathcal{K}}{D^2}\right)\right)=0
\end{eqnarray}
Solving for the stationary configuration corresponds to solving for the shape of the membrane for a given stationary velocity configuration. In addition, $\beta$ is a perturbative parameter and hence we will solve for the above equation perturbatively correcting over the pure Einstein-Gravity solution (i.e. $\beta=0$). Let the trace of the extrinsic curvature be given by
\begin{equation}\label{pert_stat_ansatz}
\mathcal{K}=\mathcal{K}_0+\beta \mathcal{K}_1
\end{equation}
%Then the above equation has a solution which can be written as 
%$$\gamma\left(\frac{1}{\mathcal{K}}+\frac{\beta \mathcal{K}}{D^2}\right)=\frac{\alpha_0-\beta \alpha_1}{D}$$
When the ansatz solution \eqref{pert_stat_ansatz} is substituted in the above equation we get the following solution for the trace of the extrinsic curvature of the membrane
\begin{equation}
\mathcal{K}=\frac{\gamma D}{\alpha_0}+\beta \left(\frac{\gamma \alpha_1 D}{\alpha_0^2}+\frac{\gamma^3 D}{\alpha_0^3}\right)
\end{equation}
where, $\alpha_0$ and $\alpha_1$ are constants which parametrise the temperature of the dual black hole \footnote{$\alpha_0$ parametrises the temperature of the black hole in absence of GB correction and $\alpha_1$ parametrises the correction of the temperature of the black hole with the addition of the GB term}. In the $\beta=0$ limit the above solution clearly matches with the stationary membrane solutions of \cite{Dandekar:2017aiv, Mandlik:2018wnw}.  
The static limit of the above solutions is when $\gamma=1$\footnote{static membrane embedded in flat spacetime $ds^2=-dt2+dr^2+r^2d\omega^2$ has killing vector $k=\partial_t$ and hence $\gamma=1$}. hence the modified soap bubble equation for a membrane dual to a static black hole in EGB gravity is
$$\mathcal{K}=\frac{ D}{\alpha_0}\left(1+\beta\frac{1}{\alpha_0^2}\right)+\beta \frac{\alpha_1 D}{\alpha_0^2}=constant$$
which matches with the observation in equation $(3.22)$ of \cite{Chen:2017hwm}
\section{Black Strings and Spectrum of Linearised Fluctuations about them}
In this section we will analyse the spectrum of linearised fluctuations about membrane configuration dual to black strings following closely the analysis of section $2.1$ of \cite{Dandekar:2016jrp}. Let us consider the metric of flat spacetime given in the form
\begin{equation}
ds^2=-dt^2+dr^2+dx^2+r^2d\Omega_n^2
\end{equation}
In these coordinates the static black string configuration is given by the surface and velocity configuration
\begin{equation}
r=1\quad \text{and}\quad u=\partial_t
\end{equation}
It is easy to check that the above configuration satisfies the membrane equation for static configurations derived above namely $\mathcal{K}=$constant. We would like to study the spectrum of linearised fluctuations which vary only along the time and the black string direction, which can be suitably parametrised by
\begin{eqnarray}
&&r=1+\epsilon\delta r(t,x)\nonumber\\
&&u=-dt+\epsilon \delta u_t(t,x)dt+\epsilon\delta u_x(t,x)dx
\end{eqnarray}
Following the analysis of similar computations undertaken in \cite{Dandekar:2016jrp} we get the frequencies of the fluctuations to be
\begin{equation}
w_1=(1-\beta)\left(\frac{-i k^2}{n}-\frac{i k}{\sqrt{n}}\right) \quad \text{and}\quad w_2=(1-\beta)\left(\frac{-i k^2}{n}+\frac{i k}{\sqrt{n}}\right)
\end{equation}
As in the case of two derivative gravity we find that there is an instability associated with the fluctuations with eigenfrequencies $w_2$. This corresponds to the Gregory-Laflamme like IR instability for EGB gravity. Following \cite{Dandekar:2016jrp} if we rescale the spatial derivatives along the black string direction such that $k=\tilde{k}\sqrt{n}$ and $\tilde{k}$ is order $\mathcal{O}(1)$ then the above frequencies become
\begin{equation}
w_1=(1-\beta)\left(-i \tilde{k}^2-i\tilde{ k}\right) \quad \text{and}\quad w_2=(1-\beta)\left(-i \tilde{k}^2+i\tilde{ k}\right)
\end{equation}
The above spectrum is in perfect agreement with the spectrum obtained in equation 3.3 of \cite{Chen:2017rxa}. The method used there were suited to studying effective black string equations with dynamics resolving distances of the order of $\mathcal{O}(\frac{1}{\sqrt{D}})$ about the black brane configuration. We see that the IR instability is resolved for wave numbers $\tilde{k}>1$. 

To study the end-point of the IR instability it is necessary to perform a full non-linear analysis as in \cite{Emparan:2015gva, Chen:2017rxa}. To perform the analysis of the non-linear $1/D$ amplitude fluctuations about the black brane configurations which resolve distances of the order of $\mathcal{O}(1/\sqrt{D})$ from the membrane perspective requires knowledge of the second order membrane equations of motion\footnote{Following arguments mentioned in \cite{Dandekar:2016jrp} it is important to take care of terms in second order equations of motion which may contribute to the black brane dynamics at higher resolution.}. Hence, in the absence of the second order membrane equations in presence of Gauss-Bonnet gravity this analysis is not carried out here.

 \section{Summary and Discussions}
  In this paper we have worked out the large $D$ membrane paradigm for Einstein-Gauss Bonnet Gravity to the leading non-trivial order in $1/D$. We have worked out the leading corrections to the black hole metric of \eqref{ansatz_metric} which solves the EGB equations to first subleading order. In the process we have also worked out the leading order constraints that needs to be put on the shape and velocity data of the membrane to get regular solutions to the EGB equations. We call these the membrane equations \eqref{scalar_mem_eq_S}, \eqref{vec_mem_eq_S}. Our expressions for metric correction and the membrane equations are valid to linear order in the Gauss-Bonnet parameter only.  
  
 We have also written down an expression for a world-volume stress tensor for the membrane based on similarity of the current system with the large $D$ black holes of Einstein-Maxwel system. The equation of conservation of this stress tensor \eqref{mem_stress_tensor} correctly reproduces the  equation of motion of the membrane. We have also worked out the spectra of light quasi-normal modes of static spherically symmetric black holes in EGB gravity in the leading large $D$ limit.  In addition we have used the modified membrane equations to derive the effective equations of static and stationary membranes. The spectrum of frequency of Gregory-Laflamme type black string instability has also been analysed in presence of a GB term using the membrane equations.
 
 In this paper we lay the groundwork towards the grander goal of understanding the status of second law of black hole thermodynamics for a general consistent classical theory of gravity with an action containing more than two derivatives. The membrane equations in the two derivative Einstein gravity at the first subleading order in $1/D$ was shown to be equivalent to the second law of black hole thermodynamics \cite{Dandekar:2016fvw, Bhattacharyya:2016nhn}. We expect that the correct interpretation of the membrane equations for EGB gravity at first subleading order in $1/D$ may give us important insight on the status of the second law of black hole thermodynamics for these theories (atleast in the large $D$ limit). The results of the current paper can be used as the base to compute the corrections to the membrane equations and the metric corrections to the first subleading order in $1/D$. 
 
The second order membrane equations can also be used to analyse the end-points of Gregory-Laflamme instability in its full non-linear glory. This will be important to understand how the membrane result matches with the analysis of effective non-uniform black string dynamics of \cite{Chen:2017rxa}. Also, it will help us better understand the equivalence between the various formalisms to obtain effective equations for black hole dynamics in large $D$. In the process a greater insight into the mechanism of the Gregory-Laflamme instability in general theories of classical gravity will be obtained..
 
 One can also work out the membrane paradigm in some other higher derivative terms in the Lovelock theories of gravity, so that some general understanding of the membrane equation of motion and stress tensor can be obtained for these theories. This will also be important to in gaining some insight on the mechanism by which second law of black hole thermodynamics may be satisfied in general consistent theories of classical gravity.  
 
In addition adapting the algorithm spelled out in \cite{Bhattacharyya:2016nhn} for higer derivative theories of gravity, one can figure out the space-time stress tensor localised on the membrane which is compatible with the linearised gravitational field at infinity from the black holes of EGB gravity. This stress tensor should also have the additional property that its conservation equation should correctly reproduce the equation of motions of the membrane. The relevant part of this stress tensor when evaluated on the world volume of the membrane should be equivalent to the expression of world-volume stress tensor given in \eqref{mem_stress_tensor}. Also, a proposal for consistent membrane stress tensor at finite $D$ which reduces to the correct stress tensor in large $D$ as in \cite{Dandekar:2017aiv} for EGB gravity will also be interesting.

 \section{Acknowledgements}
 We would like to thank N. Bannerjee, S. Bhattacharyya, Y. Dandekar, A. Kar, T. Mandal, S. Minwalla, J. Sonner and B. Withers for many useful discussions. We would also like to thank R. Emparan, S. Minwalla, J. Sonner and B. Withers for many useful comments on the preliminary draft of this paper. We would also like to thank the organisers of the workshop ``Strings Attached" at IIT Kanpur where the author presented the preliminary results.~The work of the author is supported by Ambizione grant no. $PZ00P2\_174225/1$ of the Swiss National Science Foundation (SNSF) and partially by the NCCR grant no. $51NF40-141869$ ``The Mathematics of Physics'' (SwissMAP).

\bibliographystyle{JHEP}
\bibliography{ssbib}
\end{document}